\documentclass[%
 reprint,
superscriptaddress,
nofootinbib,
nobibnotes,
 amsmath,amssymb,
 aps,
]{revtex4-1}

\usepackage{graphicx}% Include figure files
\usepackage{dcolumn}% Align table columns on decimal point
\usepackage{subfigure}
\usepackage{multirow}
\usepackage{color}

\def\bh{{\mathbb{H}}}
\def\br{{\mathbb{R}}}
\def\bb{{\mathbb{B}}}

\def\bp{{\mathbb{P}}}

\def\pd{{p_{\mathbf{d}}}}
\def\pdp{{p_{\mathbf{d'}}}}
\def\bd{{{\mathbf{d}}}}
\def\bdp{{{\mathbf{d'}}}}

\def\bm{{{\mathbf{m}}}}

\newcommand{\dws}{d_{ws}}
\newcommand{\nwt}{n_{wt}}

\newcommand{\drs}{d_{rs}}
\newcommand{\nrt}{n_{rt}}

\newcommand{\dbs}{d_{bs}}
\newcommand{\nbt}{n_{bt}}

\newcommand{\drsp}{d'_{rs}}
\newcommand{\nrtp}{n'_{rt}}

\newcommand{\dbsp}{d'_{bs}}
\newcommand{\nbtp}{n'_{bt}}

\newcommand{\mrs}{m_{rs}}

\newcommand{\mrtOne}{m_{rt1}}
\newcommand{\mrtTwo}{m_{rt2}}
\newcommand{\mbs}{m_{bs}}

\newcommand{\mbtOne}{m_{bt1}}
\newcommand{\mbtTwo}{m_{bt2}}

\newcommand{\hrsx}{h_{rs}(x)}

\newcommand{\hrtx}{h_{rt}(x)}

\newcommand{\hbsx}{h_{bs}(x)}

\newcommand{\hbtx}{h_{bt}(x)}

\newcommand{\hrsOne}{h_{rs}(1)}
\newcommand{\hrtOne}{h_{rt}(1)}
\newcommand{\hbsOne}{h_{bs}(1)}
\newcommand{\hbtOne}{h_{bt}(1)}

\newcommand{\lsb}{\left(}
\newcommand{\rsb}{\right)}
\newcommand{\lmb}{\left[}
\newcommand{\rmb}{\right]}

\newcommand{\RrsFull}{\rho_{\bd, rs}}
\newcommand{\RrtFull}{\rho_{\bd, rt}}

\newcommand{\RbsFull}{\rho_{\bd, bs}}
\newcommand{\RbtFull}{\rho_{\bd, bt}}

\newcommand{\RrsFullp}{\rho_{\bdp, rs}}
\newcommand{\RrtFullp}{\rho_{\bdp, rt}}

\newcommand{\qrsl}{q_{rs, \ell}}
\newcommand{\qrsln}{q_{rs, \ell+1}}
\newcommand{\qbsl}{q_{bs, \ell}}

\newcommand{\qrslInf}{q_{rs, \infty}}

\newcommand{\qbslInf}{q_{bs, \infty}}

\newcommand{\qrtlOneInf}{q_{rt1, \infty}}
\newcommand{\qrtlTwoInf}{q_{rt2, \infty}}
\newcommand{\qbtlOneInf}{q_{bt1, \infty}}
\newcommand{\qbtlTwoInf}{q_{bt2, \infty}}

\newcommand{\qrtlOne}{q_{rt1, \ell}}
\newcommand{\qrtlTwo}{q_{rt2, \ell}}
\newcommand{\qbtlOne}{q_{bt1, \ell}}
\newcommand{\qbtlTwo}{q_{bt2, \ell}}
\newcommand{\qrtlnOne}{q_{rt1, \ell+1}}
\newcommand{\qrtlnTwo}{q_{rt2, \ell+1}}

\begin{document}

\preprint{APS/123-QED}

%\title{Complex Contagions in Clustered Multiplex Networks}% Force line breaks with \\
\title{Clustering determines the dynamics of complex contagions in multiplex networks}

\author{Yong Zhuang}
% \altaffiliation[Also at ]{Department of ECE and CyLab, Carnegie Mellon University, Pittsburgh, PA 15213 USA.}%Lines break automatically or can be forced with \\
%\author{}%
\affiliation{
 Department of ECE and CyLab, Carnegie Mellon University, Pittsburgh, PA 15213 USA
}
\author{Alex Arenas}
\affiliation{
Departament d'Enginyeria Inform\'atica i Matem\'atiques, Universitat Rovira i Virgili, 43007 Tarragona, Spain
}
\author{Osman Ya\u{g}an}
\affiliation{
 Department of ECE and CyLab, Carnegie Mellon University, Pittsburgh, PA 15213 USA
}

\date{\today}% It is always \today, today,

\begin{abstract}
We present the mathematical analysis of generalized complex contagions in clustered multiplex networks for susceptible-infected-recovered (SIR)-like dynamics. The model is intended to understand diffusion of influence, or any other spreading process implying a threshold dynamics, in setups of interconnected networks with significant clustering. 
The contagion is assumed to be general enough to account for a content-dependent linear threshold model, where each link type has a different weight (for spreading influence) that may depend on the content (e.g., product, rumor, political view) that is being spread.
Using the generating functions formalism, we determine the conditions, probability, and expected size of the emergent {\em global} cascades. This analysis provides a generalization of previous approaches and is specially useful in problems related to spreading and percolation.
The results present non trivial dependencies between the clustering coefficient of the networks and its average degree. In particular, several phase transitions are shown to occur depending on these descriptors.
Generally speaking, our findings reveal that increasing clustering decreases the probability of having global cascades and their size, however this tendency changes with the average degree. There exists a certain average degree from which on clustering favours the probability and size of the contagion. 
By comparing the dynamics of complex contagions over multiplex networks and their monoplex projections, we demonstrate that  ignoring link types and aggregating network layers may lead to inaccurate conclusions about contagion dynamics, particularly when the correlation of degrees between layers is high.

%, i.e., cases where influence starts from a single individual and reaches a positive fraction of the population.
%We study the diffusion of influence, i.e., complex contagions, in \textit{clustered multiplex} networks.
%The contagion is assumed to take place
%according to the content-dependent linear threshold model proposed in [Ya\u{g}an and Gligor, PRE $\mathbf{86}$, 036103 (2012)], where each link type in the network has a
%different weight (for spreading influence) that may depend on the content (e.g., product, rumor, political view) that is being spread.
%We determine the conditions, probability, and expected size of {\em global} cascades, i.e., cases where influence starts from a single individual and reaches a positive fraction of the population. We show that 
%clustering decreases the probability and size of cascades when average degree in the network is small, whereas after a certain value of the average degree, clustering is shown to facilitate cascades. 
%By comparing the dynamics of complex contagions over multiplex networks and their monoplex projections, we demonstrate that  ignoring link types and aggregating network layers may lead to inaccurate conclusions about contagion dynamics, particularly when assortativity is high. Finally, we show for the first time that linear threshold models do not necessarily exhibit two phase transitions as previously reported. In fact, depending on the assortativity, it is possible to observe four phase transitions both in monoplex and multiplex cases (with two link types).

\begin{description}
\item[PACS numbers]
\end{description}
\end{abstract}

\pacs{Valid PACS appear here}% PACS, the Physics and Astronomy
\maketitle

%\tableofcontents

\section{Introduction}
\label{sec:introduction}
%With the development of network systems, today's worldwide complex systems ranging from telecommunication systems and transportation networks to human society consist of different types of links.
%For example, individuals in a society will be connected with each other through different relationships, e.g., colleagueship, kinship, companionship, etc..
The study of dynamical processes on real-world complex networks has been an active research area over the past decade.
Some of the most widely studied problems include
cascading failures in interdependent networks \cite{Buldyrev,Vespignani,YaganQianZhangCochranLong,brummitt2012suppressing,radicchi2013abrupt,reis2014avoiding},  
{\em simple} contagions (e.g., diffusion of information, disease propagation in human and animal populations \cite{albert2000error, cohen2000resilience, cohen2002percolation, leicht2009percolation, nagler2011impact, Murray, son2011percolation,baxter2012avalanche,dickison2012epidemics,AndersonMay, SahnehScoglioMieghem,OY13,valdez2013triple, zhao2013percolation, cellai2013percolation,azimi2014k,bianconi2014multiple, radicchi2015percolation}, etc.), and {\em complex} contagions  (e.g., diffusion of influence, beliefs, norms, and innovations in social networks \cite{gleeson2008cascades,WattsExternal,YaganPRE}). Recently, the attention was shifted from single, isolated networks to multiplex and multi-layer networks \cite{Alex_PRX,SahnehScoglioMieghem,OY13,YaganPRE,zhuang2015information,brummitt2012multiplexity, kivela2014multilayer, Domenico10062014, de2013mathematical, min2014network, serrano2015escaping, granell2013dynamical,
baxter2014weak,
de2016physics, wu2016influence, boccaletti2014structure, cellai2016multiplex}. This shift is primarily driven by the observation that links 
 in a network might be categorized according to the nature of the relationship they represent (e.g., friends, family, office-mates) as well as according to the social network they belong (e.g., Google+ vs. Facebook links), and each link type might play a different role in the dynamical process.

In this work, we focus on the analysis of complex contagion processes that take place on multiplex (or, multi-layer) networks. In doing so, we adopt the content-dependent linear threshold model of social contagions proposed by Ya\u{g}an and Gligor \cite{YaganPRE}. Their framework is a generalization of the linear threshold model introduced by Watts \cite{WattsExternal} and is based on individuals adopting a behavior when their {\em perceived} proportion of {\em active} neighbors exceed a certain threshold; the key to that modeling framework is that one's perceived influence depends on 
the {\em types} of the relationships they have and the {\em context} in which diffusion is being considered. 
More precisely, each individual in the network can be in one of the two states, {\em active} or {\em inactive}. Each link type-$i$ is associated with a content-dependent weight $c_{i}$ in $[0, \infty]$ that encodes the relative importance of this link type in spreading the given content. Then, an inactive node with $m_i$ active neighbors and $d_i - m_i$ inactive neighbors via type-$i$ links
turns active only if  
\[
\frac{\sum_i c_i m_i}{\sum_i c_i d_i} \geq \tau
\]
where $\tau$ is the node's threshold drawn from  a distribution $P(\tau)$. Once \textit{active}, a node stays active forever. 
 %nodes will not be back to \textit{inactive}.

%Then, with a small number of \textit{active} nodes in the network, nodes update the states in discrete time steps synchronously.
%The nodes will update their states synchronously at times $t = 0, 1, \dots.$
%{\em Inactive} nodes will be activated at time $t$ if the fraction of their {\em active} neighbors exceeds their threshold at time $t - 1$.

%This process can be investigated in the context of marketing \cite{gladwell2006tipping}.
%Individuals will make decisions based on the decisions of others instead of their own information, which is referred to a herd-like behavior in \cite{WattsExternal}.

Ya\u{g}an  and Gligor analyzed \cite{YaganPRE} the content-dependent linear threshold model in multiplex networks and derived the conditions, probability, and expected size of {\em global} cascades, i.e., cases where activating a randomly selected node leads to activation of a positive fraction of the population in the limit of large system sizes.
However, their  multiplex network model was formed by combining independent layers of networks (one for each link type), where each layer is generated by the {\em configuration model}  \cite{MN01, MN03}.  
Although a good starting point, configuration model is known to generate networks that can not accurately capture some important aspects of real-world social networks, most notably the property of {\em high
clustering} \cite{WattsStrogatz,SerranoBoguna2}.
Informally known as the phenomenon that \lq\lq friends of our friends" are likely to be our friends, clustering has been shown to affect  
the dynamics of various diffusion processes \cite{MN09,
gleeson2010clustering,
hackett2011site, hackett2011cascades,huang2013robustness, colomer2014double,zhuang2015information, cozzo2015structure} significantly.

With this in mind, our main contribution in this paper is to provide a thorough analysis of influence diffusion process (a complex contagion) in \textit{clustered} multiplex networks. In particular, we study the content dependent linear threshold model in a multiplex network model where each link type (or, network layer) is formed by the clustered random network model proposed by Newman \cite{MN09} and Miller \cite{JM09}. We solve for the critical threshold, probability, and expected size of global cascades and confirm our analytical results via extensive simulations. The main observation from 
our results is that 
clustering has a double-faceted impact on the probability and expected size of global cascades. Namely, we show that 
clustering decreases the probability and size of cascades when average degree in the network is small, whereas after a certain value of average degree, clustering is shown to facilitate cascades. 

We also compare the dynamics of complex contagions over multiplex networks and their {\em monoplex} projections. There has been recent interest \cite{Alex_PRX} in understanding whether monoplex projection of a multiplex network (obtained by ignoring the colors of edges and aggregating the layers) can still capture the essential properties (e.g., cascade threshold and size) of a diffusion process. In the affirmative, this would eliminate the need for considering the full multiplex structure of real-world systems in tackling similar problems. 
We show that even in the simplest case where all link types have the same influence weight (i.e., $c_1=c_2 = \cdots$),
monoplex theory may {\em not} be able to capture contagion dynamics accurately, reinforcing the need for studying multiplex networks in its correct setup.
We demonstrate that the accuracy of monoplex theory in capturing cascade dynamics over multiplex networks depends tightly on the assortativity (i.e., correlation between the degrees of connected pairs) of the network. For instance, when assortativity is negligible, monoplex theory is seen to predict  cascade dynamics very well, while in highly assortative cases its ability to predict contagion behavior diminishes significantly.

Finally, we proof the possibility of an unforeseen behavior in the dynamics of complex contagions in multiplex networks, i.e., that of observing more than two phase transitions in the cascade size as the mean degrees in network layers increase. 
It has been reported  \cite{WattsExternal, OY13}  many  times that threshold models of complex contagion exhibit two phase transitions as the average degree increases; a second-order transition at {\em low} degrees marking the formation of a giant  component of {\em vulnerable} nodes and a first-order transition at {\em high} degrees due to increased {\em local} stability of nodes. Here, we consider a multiplex where one layer has degree distribution $\textrm{Poi}(\lambda)$ while the degree in the second layer follows $\textrm{Poi}(\lambda/\alpha)$  with probability $\alpha$ and is zero with probability $1-\alpha$. In this setting, we observe that in general there exist  two intervals of $\lambda$ for which cascades are possible, amounting to {\em four} phase transitions as opposed to two; also, it is seen that only the first transition is second-order while the remaining ones are first order. However, depending on the value of $\alpha$, these regions may overlap (with overlap starting when $\alpha$ exceeds a critical value) resulting again with only two phase transitions; see Section \ref{sec:networkMultiplePhases} for details. 

%, it is also observed that monoplex theory can give accurate predictions under certain parameter settings. 

% at least in some simple settings where 

%at least in simple cases where all link types carry an equal weight or have the same transmissibility. 

The paper is organized as follows.
We give details of the models applied in this study and the problem to be considered in Section \ref{sec:systemModels}.
In Sections \ref{sec:conditionProbabilityGlobalCascades} and \ref{sec:expectedCascadeSize} we present the main results of this work, and confirm it through extensive computer simulations in Section \ref{sec:numericalResults}.
In Section \ref{sec:comp_monoplex_multiplex}, we make a comparison between complex contagions in monoplex networks and multiplex networks, and also demonstrate the new phenomenon about the number of phase transitions.
Finally, Section \ref{sec:conclusion} summarizes our work and gives future directions.

\section{Model: structure and dynamics}
\label{sec:systemModels}

%In this section, we will introduce the models considered in this paper and the problem of interest.
%To put it simply

\subsection{Random Graphs with Clustering}
\label{sec:networkClustering}

%Namely, consider a vertex set $V = {1, 2, \dots, n}$, where each vertex is independently assigned a random number of {\em stubs} according to a probability distribution $\{p_k\}_{k=0}^{\infty}$; i.e., the degree $d_i$ of vertex $i$ equals $k$ with probability $p_k$ for any positive integer $k$. 
%Then, stubs are randomly paired with each other to form edges until no free stubs is left.

Our goal is to study complex contagion processes in synthetic networks that capture some important aspects of real-world networks but  otherwise are generated {\em randomly}. 
It is known \cite{MN01,MN03} that the widely used configuration model \cite{MN01} generates tree-like graphs with number of cycles approaching to zero as the number of nodes gets large. However, most social networks exhibit {\em high clustering}, informally known as the propensity of a \lq\lq friend of a friend" to be one's friend. Put differently, real-world social networks are usually not tree-like and instead have considerable number of cycles, particularly of size three; i.e., {\em triangles}.
With this in mind, Miller \cite{MM95}
and Newman \cite{MN03} proposed a modification on the configuration model to enable generating random graphs with given degree distributions and tunable clustering. 

%We generate random graphs with clustering using the \textit{generalized configuration model} introduced by Newman 
%\cite{MN09}
%and Miller \cite{JM09}.
The model proposed in \cite{MM95, MN03} is often referred to as random networks with clustering and is based on the following algorithm.
%Consider a joint degree distribution $\{p_{st}\}_{s,t=0}^{\infty}$ that gives the probability that a node has $s$ single edges and $t$ triangles; e.g., see node $2$ in Figure \ref{fig:clusteringA} that has two single edges and one triangle.
Given a joint degree distribution $\{p_{st}\}_{s,t=0}^{\infty}$ that gives the probability that a node has $s$ single edges and $t$ triangles, each node will be given $s$ stubs labeled as {\em single} and $2t$ stubs labeled as {\em triangles} with probability $p_{st}$, for any $s,t=0, 1,\ldots$. Then, stubs that are labeled as single are {\em randomly} joined to form single edges that are not part of a triangle, whereas {\em pairs} of triangle stubs from three nodes are {\em randomly} matched to form triangles between the three participating nodes; the total degree of a node is then distributed by $p_k = \sum_{s,t: s+2t = k} p_{st}$.
As in the standard configuration model, it can be shown that the number of cycles formed by single edges goes to zero as $n$ gets large, and so does the number of cycles of length larger than three \cite{MN01}.

We quantify the level of clustering using the widely recognized \textit{global} clustering coefficient \cite{MN03},
%and \textit{local} clustering coefficient \cite{AB00}.
defined via
\begin{align}\nonumber
    C_{\textrm{global}} = \frac{3 \times (\text{number of triangles in network})}{\text{number of connected triples}}.
\end{align}
Here, \lq\lq connected triples'' means a single vertex connected by edges to two others. 
%On the other hand, the local clustering is defined as the average
%\begin{align}
%    C_{\textrm{local}} = \frac{1}{n^{*}}\sum_{i}C_{i},
%\end{align}
%where $C_i$ denotes the clustering coefficient for node $i$ %given by
%\begin{align}
%    &C_{i} = \frac{\text{number of triangles connected to vertex }i}{\text{number of connected triples centered on vertex }i}.
%    \label{eq:local_clustering_defn}
%\end{align}
%Here, $n^{*}$ is the number of nodes whose $C_{i}$ is well-defined in the network; i.e., number of nodes where the denominator at (\ref{eq:local_clustering_defn}) is nonzero.
It was shown in \cite{MN01} that  $C_{\textrm{global}}$ is {\em positive} in the random clustered network model, while it  approaches to zero with increasing network size in the standard configuration model.

\begin{figure*}[!t]
    \centering
    \subfigure[]{
    \hspace{0cm}
    \includegraphics[width=0.49\textwidth] {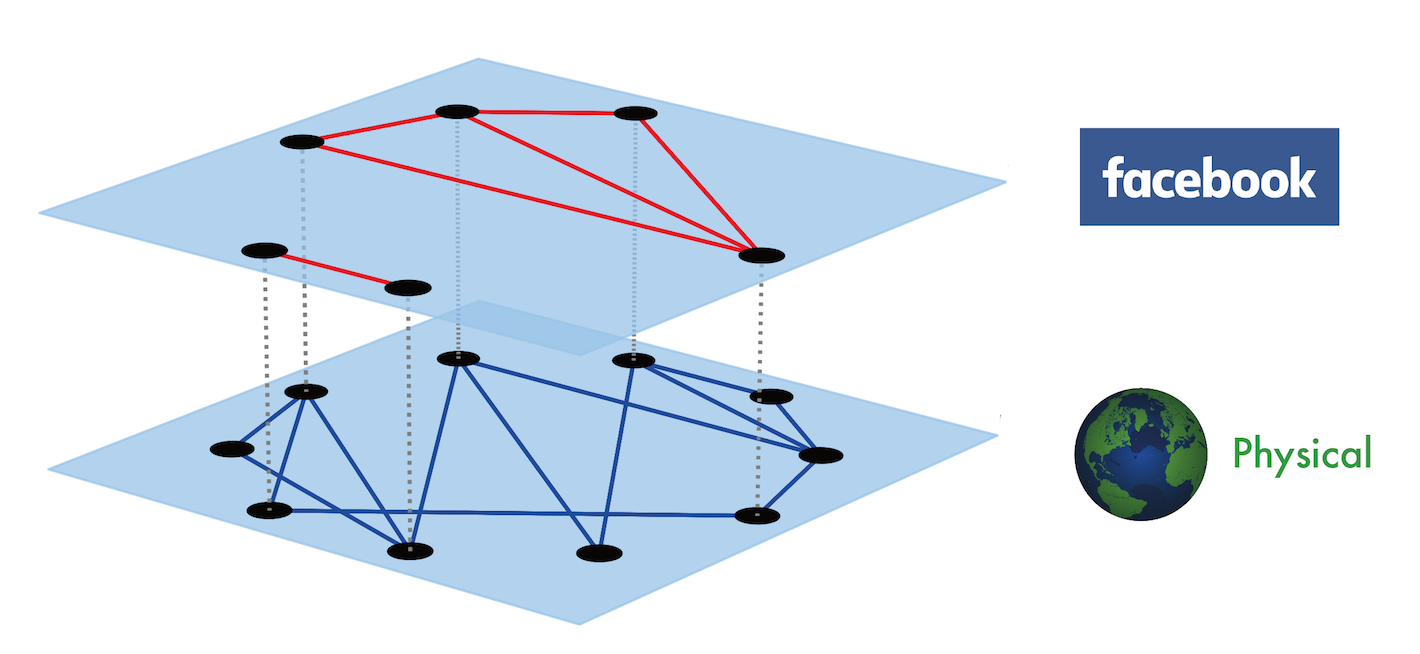}
    \label{fig:expmonoplexVSMultiplexLimitedAssortativity01}
    }
    \hspace{2em}
    \subfigure[]{\hspace{0cm} \includegraphics[width=0.4\textwidth] {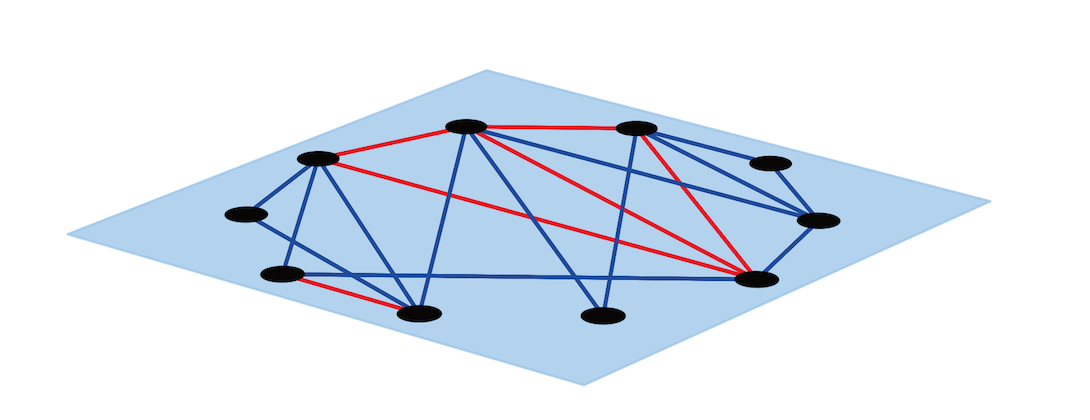} 
    \label{fig:expmonoplexVSMultiplexLimitedAssortativity099}
    }
    \caption{\sl Illustration of multi-layer and multiplex network representations of our model.
    In (a), we see a multilayer network (e.g., a Physical communication layer and a Facebook layer) with overlapping vertex sets; vertical dashed lines represent nodes corresponding to the same individual.
   % The upper-layer network represents the relationships in Facebook network, while the lower-layer network represents the physical network.
    In (b), we see the equivalent representation of this model by a multiplex network.
    Edges from Facebook are shown in red and edges from the physical network are shown in blue.
    }
\label{fig:multiplexNetwork}
\end{figure*}

\subsection{Multi-layer and Multiplex Network Models}
\label{sec:multiplexNetworkModelsWithClustering}
In this paper, we consider a multiplex network where links are classified into different types, {\em or} colors.
%To make it easier for understanding, we define the model in context of an colored degree-driven random graphs proposed in \cite{soderberg2003random, soderberg2002general}.
For ease of exposition, we consider the case with only two colors, {\em red} and {\em blue}, but the discussion can easily be extended to arbitrary number of colors.
%Simply put, we let $\bw$ and $\bf$ denote the two constituent layers of networks with the possible motivation that $\bw$ models the \emph{physical} contact network among individuals, i.e., models face-to-face relationships, while network.
Let $\br$ and $\bb$ denote the sub-networks formed by red and blue edges, respectively. 
A possible motivation is that $\br$ models the \emph{kinship} contact network among individuals, while the network $\bb$ stands for the colleagueship network.
Alternatively, we can think of $\bb$ modeling the physical (e.g., face-to-face) relationships among human beings while $\br$ models connections through an online social network (e.g., Facebook).

In line with the second motivation, we assume that network $\bb$ is defined on the vertices $\mathcal{N} = \{1, \dots, n\}$,  while $\br$ contains only a {\em subset} of the nodes in $\mathcal{N}$ to account for the fact that not every individual participates in online social networks; e.g., we assume that each vertex in $\mathcal{N}$ participates in $\br$ independently with probability $\alpha \in (0, 1]$, meaning that the set of vertices of $\br$ constitutes $\alpha$-fraction of the whole population. We illustrate in
Figure \ref{fig:multiplexNetwork} two equivalent representations of this model, first shown as a multi-layer network with overlapping vertex sets, and second as a multiplex network.
%with mult an illustration of an ensemble of the colored degree-driven random graphs.

%\begin{figure*}[!ht]
%	\centering
%    \includegraphics[width=0.6\textwidth]{figs/multilayerNetwork_v2.png}
%    \vspace{-16mm}
%    \caption{\sl An illustration of a multiplex network with three relationships.
%    Edges in red, green, and blue are classified into 3 types of relationship (e.g., friends in physical network, in Twitter, and in Instagram).
%        Besides, two nodes may be connected by multiple relationships rather than only one.
%        For example, the yellow edge represent two individuals in the networks are connected by two relationship (e.g., red and green), while the pink one represents that there are three relationships between the two nodes.
%    }
%	\label{fig:multiplexNetwork}
%\end{figure*}

%The case where $|\mathcal{N}_{\br}| = o(n)$ has been considered in \cite{OY13} and it was shown that most properties pertaining to the propagation information are unaffected by the existence of the upper layer $\br$; i.e., when the online social network has a negligible size compared to the whole population, it does not impact the threshold or size of information epidemics.

We generate both $\br$ and $\bb$ from the generalized configuration model 
described in Section \ref{sec:networkClustering}; i.e., both are random networks with clustering.
In particular, we let $\{p_{st}^{r}, s, t = 0, 1, \dots\}$ and $\{p_{st}^{b}, s, t = 0, 1, \dots\}$ denote the joint distributions for single edges and triangles for $\br$ and $\bb$, respectively.
Then both networks are generated independently according to the algorithm described in Section \ref{sec:networkClustering}, and they are denoted respectively by $\br = \br(n;\alpha, p_{st}^{r})$ and $\bb = \bb(n;p_{st}^{b})$.
We define the overall network $\bh$ over which influence spreads as the {\em disjoint} union $\bh = \br \coprod \bb$ and represent it by $\bh(n;\alpha, p_{st}^{r}, p_{st}^{b})$.
Here, the disjoint union operation implies that we still distinguish $\br$-edges from $\bb$-edges in $\bh$, meaning that  it is a {\em multiplex} network. 
%With this distinguishment, we can assign different content parameters for each link type.
%this is done to accommodate the
%possibly different rates (or, even rules) of information propagation across the two networks.
%To this end, an equivalent representation of $\bh$ would be a {\em multiplex} network with different types (or, colors) of edges.

%With these definitions in mind, we use the colored degree-driven networks.

%Let $\drs$ and $\dbs$ denote the random variables corresponding to the number of \textit{single edges} for a node in $\br$ and $\bb$, respectively, while $\nrt$ and $\nbt$ are defined similarly for the number of triangles assigned; i.e., the degree of a node from triangle {\em edges} in $\br$ is given by $d_{rt}=2n_{rt}$ and similarly for $d_{bt}$.

We denote the {\em colored} degree $\mathbf{d}$ of a node in $\bh$ by
\begin{align}
    \mathbf{d} = (d_{rs}, 2n_{rt}, d_{bs}, 2n_{bt})
\end{align}
meaning that it has $d_{rs}$ \textit{single edges} and $2n_{rt}$ \textit{triangle edges} in network $\br$, and $d_{bs}$ \textit{single edges} and $2n_{bt}$ \textit{triangle edges} in network $\bb$; here $\nrt$ and $\nbt$ are defined as the number of {\em triangles} assigned to this node in $\br$ and $\bb$, respectively. The distribution of this colored degree is denoted by $p_{\mathbf{d}}$
and can be computed directly from $p_{st}^{r}$, $p_{st}^{b}$ and $\alpha$.

%Under the assumptions enforced here, the distribution of this colored degree is given by
%{\color{red} I do not understand the next formula. Can we write it as $\alpha p_{red} + (1-\alpha) p_{blue}$ or similar. the notation is highly complex, many indices make the reader the lose attention, discomfort and misunderstanding. Before continuing we have to prescribe a simple but meaningful notation through the rest pf the paper. BTW, is the last parenthesis in the equation ok?}
%\begin{align}
 %   p_{\mathbf{d}} \hspace{-.4mm}= \left(\alpha p^r_{d_{rs}n_{rt}}  \hspace{-.4mm}+ \hspace{-.4mm} (1 - \alpha) \mathbf{1}[d_{rs} = 0 \wedge n_{rt} = 0]\right)p^b_{d_{bs}n_{bt}}
   % \label{eq:possibilityPd}
  % \notag
%\end{align}
%where the term $ (1 - \alpha) \mathbf{1}[d_{rs} = 0 ~\wedge~ n_{rt} = 0]$ accounts for the fact that if the node does not belong to $\br$ (which happens with probability $1-\alpha$), then its degree from red-single and red-triangle edges will both be zero.
%{\color{red} This explanation helps, but is not usual, we have to find a way to put this as a continuous function with no conjunctions. If the node is not red, is blue, right? Then why not writing with probability $(1-\alpha)$ is blue?}

\begin{figure*}[!t]
    %\centering
    \subfigure[]{\hspace{\fill} \includegraphics[totalheight=0.11\textheight,width=0.06\textwidth] {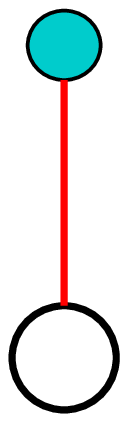}}
    \hspace{4em}
    \subfigure[]{\hspace{\fill} \includegraphics[totalheight=0.11\textheight,width=0.16\textwidth] {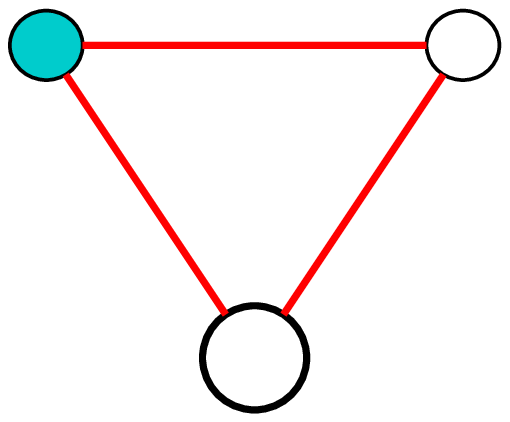}}
    \hspace{2em}
    \subfigure[]{\hspace{\fill} \includegraphics[totalheight=0.11\textheight,width=0.16\textwidth] {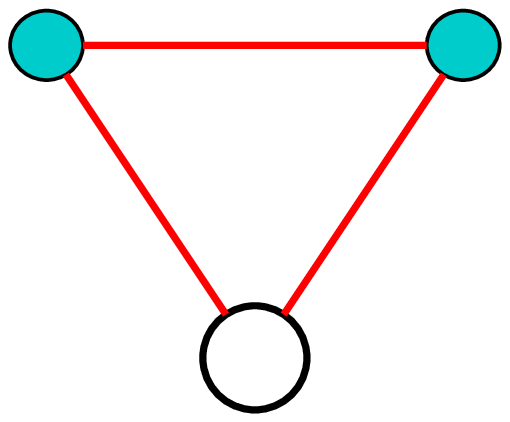}}
    \caption{\sl Illustration of three cases that would be counted as $\mrs$, $\mrtOne$, and $\mrtTwo$, respectively for the number of active nodes.
    Nodes shown in filled (green) circles are \textit{active} while those shown in non-filled circles  are \textit{inactive}.
    }
\label{fig:differentTypeActiveEdge}
\end{figure*}

\subsection{Content-dependent Linear Threshold Model for Social Contagion}
\label{sec:linearThresholdModel}
The classical linear threshold model by Watts \cite{WattsExternal} is based
on indviduals adopting a behavior when the fraction of their {\em active} neighbors exceed a certain threshold. Namely, 
an inactive node $i$ with $m_i$ active neighbors and $d_i - m_i$ inactive neighbors will become active only if $m_i/d_i$ exceeds $\tau_i$ drawn from a distribution $P(\tau)$;
once active, a node can not be deactivated. 
%threshold model , we apply the linear threshold model proposed in \cite{YaganPRE}.
%In Watts'\ threshold model
A major concern with this model is that it assumes all links in the network have the same importance, irrespective of the context that the spreading is being considered.
However, in real world contagion processes, 
each link type (e.g., co-workership vs.~family or physical links vs.~online social network links) may play a different role in different
cascade processes. For example, in the spread of a new consumer product amongst the population, 
a video game would be more likely to be promoted among high school classmates 
rather than among family members; the situation 
would be exactly the opposite
in the case of a new cleaning product \cite{TangYuanMaoLiChenDai}.
% In a similar way, one can make a case for 
%distinguishing between links in different online social networks (e.g., Google+, Facebook, Twitter, etc.),
%However, in real world contagion processes, the link types in network (e.g., colleagueship, kinship, companionship, etc.), might have a different impact in the spreading and the relative importance of  each type should have different weights on the influence diffusion process.

To address the aforementioned drawbacks, Ya\u{g}an and Gligor \cite{YaganPRE} proposed a content-dependent linear threshold model for social contagion in multiplex networks. In this model, each link type is associated with a content dependent parameter $c_{i}$ in $[0, \infty]$ that measures the relative bias type-$i$ links have in spreading the content.
Then, an inactive node with $m_i$ active neighbors and $d_i - m_i$ inactive neighbors through link type-$i$ will turn {\em active} if $\frac{\sum_i c_i m_i}{\sum_i c_i d_i} \geq \tau$.

In this work, we will analyze complex contagions in $\bh$ 
under the content-dependent threshold model introduced in \cite{YaganPRE}.
Consider a node with colored degree $\bd = \lsb \drs, 2\nrt, \dbs, 2\nbt\rsb$ and
 {\em active}-degree %$\bm$; i.e., with
\[
\bm = \lsb \mrs, \mrtOne, \mrtTwo, \mbs, \mbtOne, \mbtTwo \rsb,
\]
%to denote its number of three different type of {\em active} neighbors connected through specific edges at time $t - 1$ in Figure \ref{fig:differentTypeActiveEdge}.
where $\mrs$ (resp. $\mbs$) gives the number of {\em active} neighbors connected through {\em red single} edges (resp. {\em blue single} edges), and 
$\mrtOne$ and $\mrtTwo$ (resp. $\mbtOne$ and $\mbtTwo$) give the number of red (resp.~blue) triangles  with one and two active neighbors, respectively; see Figure \ref{fig:differentTypeActiveEdge} for demonstration of three cases counted as  $\mrs$, $\mrtOne$, and $\mrtTwo$, respectively.
Next, for a given content to spread over $\bh$, let $c_r$ and $c_b$ denote the weight of $red$- and $blue$-edges, respectively, in spreading this content. Without loss of generality, we set $c := \frac{c_{r}}{c_{b}}$.
Then, the probability that an inactive node with degree $\bd$ and 
active-degree $\bm$ turns active is
given by
\begin{align}
&    F(\bm, \bd) 
    \label{eq:responseFunction}
 \\
& = \bp\lmb\frac{\hspace{-.5mm}c\lsb \mrs \hspace{-.5mm}+\hspace{-.5mm} \mrtOne \hspace{-.5mm}+\hspace{-.5mm} 2\mrtTwo\rsb \hspace{-.5mm} + \mbs + \mbtOne \hspace{-.5mm}+ \hspace{-.5mm} 2\mbtTwo}{c\lsb \drs + 2\nrt \rsb + \dbs + 2\nbt}  \ge \tau \rmb
\nonumber
\end{align}
Hereafter, the function $F(\bm, \bd)$ will be referred to as the \textit{neighborhood response function} \cite{DoddsWatts, hackett2011cascades}.

%{
%    \footnotesize
%\begin{align}
%    \bp\lmb\frac{c_1\lsb \mrs + \mrtOne + 2\mrtTwo\rsb + c_2 \lsb \mbs + \mbtOne + 2\mbtTwo \rsb}{c_{1}\lsb\drs + \nrt\rsb + c_{2}\lsb \dbs + 2\nbt \rsb}\rmb := F(\bm, \bd).
%\end{align}
%}

%\begin{widetext}
%\begin{align}
%    \bp\lmb\frac{c_r\lsb \mrs + \mrtOne + 2\mrtTwo\rsb + c_b\lsb\mbs + \mbtOne + 2\mbtTwo\rsb}{c_r\lsb \drs + 2\nrt \rsb + c_b \lsb\dbs + 2\nbt\rsb}  \ge \tau \rmb := F(\bm, \bd).
%    \label{eq:activeProbability}
%\end{align}
%\end{widetext}

\subsection{The Problem}
We consider the  diffusion of influence over $\bh$ that is initiated by a randomly selected node.
Our main goal is to derive the conditions, probability, and expected size of {\em global} cascades, i.e., cases where influence starts from a single individual and reaches a positive fraction of the population
in the large $n$ limit. Of particular interest will be to reveal the effect of clustering coefficient $C_{\textrm{global}}$ and content parameter $c$ on these quantities.

%In the following experiments, we will use two interesting metrics, the probability to trigger a global cascade and the expected size of it, to analyze it.
%In addition, we will see how clustering and content parameter affect them.

%\section{Main Results}
%\label{sec:mainResults}

\section{Condition and Probability of Global Cascades}
\label{sec:conditionProbabilityGlobalCascades}

In this section, we derive the condition and probability of global cascades in clustered multiplex networks; expected size of global cascades is handled separately in
Section \ref{sec:expectedCascadeSize}.
As mentioned in Section \ref{sec:multiplexNetworkModelsWithClustering}, we restrict our attention to networks with only two link types, labeled as {\em blue} and {\em red} edges, respectively. 
Distinguishing further the edges based on whether or not they are part of a triangle, we obtain four types of edges in our clustered multiplex network model, labeled as \textit{red single} edges, \textit{red triangles edges}, \textit{blue single} edges, and \textit{blue triangle edges}; these are denoted by $rs-$, $rt-$, $bs-$, and $bt-$, respectively.

To analyze the influence diffusion process, we consider a branching process \cite{athreya2012branching} that starts by activating a node selected randomly from among all nodes.
Starting with the neighbors of the seed node, we explore and identify all nodes that are reached and activated, continuing recursively until the branching process stops.
%
%\begin{widetext}
%\begin{align}
%     \RrsFull =  F\lmb\lsb1, 0, 0, 0, 0, 0\rsb, \bd\rmb = \bbp\left[\frac{c}{c\lsb\drs + 2\nrt\rsb + \dbs + 2\nbt} \ge \tau\right]   
 %   \label{eq:FrsFull}
%    \\
%     \RrtFull = F\lmb\lsb0, 0, 1, 0, 0, 0\rsb,  \bd\rmb = \bbp\left[\frac{2c}{c\lsb\drs + 2\nrt\rsb + \dbs + 2\nbt} \ge \tau\right]
%    \label{eq:FrtFull}
%    \\
%     \RbsFull =  F\lmb\lsb0, 0, 0, 1, 0, 0\rsb, \bd\rmb = \bbp\left[\frac{1}{c\lsb\drs + 2\nrt\rsb + \dbs + 2\nbt} \ge \tau\right]
%    \label{eq:FbsFull}
%    \\
%     \RbtFull = F\lmb\lsb0, 0, 0, 0, 0, 1\rsb,  \bd\rmb = \bbp\left[\frac{2}{c\lsb\drs + 2\nrt\rsb + \dbs + 2\nbt} \ge \tau\right]
    %\label{eq:FbtFull}
%\end{align}
%
%\end{widetext}
%The mathematical definitions of $\RrsFull$, $\RrtFull$, $\RbsFull$, and $\RbtFull$ are described in (\ref{eq:FrsFull}) - (\ref{eq:FbtFull}).
%
%Here, we begin to derive the condition and probability of global cascades.
Let $H(x)$ denote the generating function \cite{HW13} for the ``\textit{finite} number of nodes that are reached and influenced'' by the branching process \cite{athreya2012branching} initiated by a randomly selected node.
We will derive an expression for $H(x)$ using $\hrsx$, $\hrtx$, $\hbsx$, and $\hbtx$, where $\hrsx$ (resp.~$\hbsx$) stands for the generating function for the ``\textit{finite} number of nodes reached by following a randomly selected \textit{red single} (resp.~\textit{blue single}) edge''; $\hrtx$ and $\hbtx$ are defined similarly for \textit{red triangle} and \textit{blue triangle} edges, respectively.
Then, $H(x)$ takes the form

\begin{align}
H(x) = x\sum_{\bd}\pd D(\drs, \nrt, \dws, \nwt),
\label{eq:Hx}
\end{align}
where 
\begin{align}
D(y, z, m, \ell) := \hrsx^{y}\hrtx^{z}\hbsx^{m}\hbtx^{\ell}
\label{eq:defn_D}
\end{align}

The validity of (\ref{eq:Hx}) can be seen as follows.
The term $x$ stands for the node that is selected randomly and set \textit{active} to initiate the cascade.
This node has a degree $\bd = \lsb \drs, 2\nrt, \dbs, 2\nbt \rsb$ with probability $\pd$.
The number of nodes reached by each of its $\drs$ (resp.~$\dbs$) \textit{red} single edges (resp.~\textit{blue} single edges) has a generating function $\hrsx$ (resp.~$\hbsx$). Considering its triangle edges in a similar manner,
 we see from the {\em powers} property of generating functions \cite{MN01} that when the initial node has degree $\bd$ the number of nodes influenced in this process has a generating function $\hrsx^{\drs}\hrtx^{\nrt}\hbsx^{\dbs}\hbtx^{\nbt}$.
Averaging over all possible degrees $\bd$ of the initial node, we get (\ref{eq:Hx}).

For (\ref{eq:Hx}) to be useful, we shall derive expressions for the generating functions $\hrsx$, $\hrtx$, $\hbsx$, and $\hbtx$.
As will  become apparent soon, there are no explicit equations defining these functions.
Instead, we should seek to derive {\em recursive} equations to define each generating function in terms of the others.
These steps are taken in the next sections where we first focus on deriving $\hrsx$ and $\hbsx$ (Section \ref{sec:influencePropagationSingle}) followed by derivations of $\hrtx$ and $\hbtx$ (Section \ref{sec:influencePropagationTriangle}).

%Before we derive recursive relations for the aforementioned generating functions, we need to call for the definition of vulnerability with respect to two different relationships.

Given that random networks with clustering are free of cycles of size larger than three, it is clear that the initial stages of the branching process will expand largely because of {\em vulnerable} nodes that can get activated either by one or two active neighbors. In our formulation, the multiplex nature of the network calls for defining the notion of the vulnerability with respect to link types as well  \cite{YaganPRE}.   
Throughout, we say that a node is $\br$-vulnerable (resp. $\bb$-vulnerable)
if it gets activated by a {\em single} active connection through a \textit{red} link (resp. \textit{blue} link).
%; clearly
%with our content-dependent threshold model, a node can be $\br$-vulnerable but not $\bb$-vulnerable, or vice versa. 
We define $\RrsFull$ and $\RbsFull$ as the probability that a node is $\br$-vulnerable and $\bb$-vulnerable, respectively.
We also define $\RrtFull$ (resp.~$\RbtFull$) as the probability that a node gets activated by having {\em two} active neighbors via \textit{red} (resp.~\textit{blue}) edges. In other words,
 $\RrtFull$ (resp.~$\RbtFull$) gives the probability that a node gets activated by having a red (resp.~blue) triangle with both neighbors being active; see Figure \ref{fig:differentTypeActiveEdge}. More precisely, we set
 $\RrsFull =  F\lmb\lsb1, 0, 0, 0, 0, 0\rsb, \bd\rmb$, 
$\RrtFull = F\lmb\lsb0, 0, 1, 0, 0, 0\rsb,  \bd\rmb$, 
$\RbsFull =  F\lmb\lsb0, 0, 0, 1, 0, 0\rsb, \bd\rmb$,
and
$\RbtFull = F\lmb\lsb0, 0, 0, 0, 0, 1\rsb,  \bd\rmb$.

%Following \cite{YaganPRE}, we say that a set of nodes that are vulnerable w.r.t. at least one of the networks form a vulnerable component if in the subgraph (of $\bh$) induced by this set of nodes, activating any node leads to the activation of all the nodes in the set.
%Besides, from the perspective of \textit{directed} graph, it relates to a \textit{strongly connected component} \cite{newman2001random, dorogovtsev2001giant}.
%The more detailed discussion can be found in \cite{YaganPRE}.

\subsection{Influence Propagation via \textit{Red Single} Edges}
\label{sec:influencePropagationSingle}

We start by deriving recursive equations for $\hrsx$ and $\hbsx$ by focusing on the number of nodes reached and influenced by following one end of a single edge in $\br$ and $\bb$, respectively.
%Because of the similarly, we only explain the derivation of $\hrsx$.
In what follows, we only derive $\hrsx$ since the computation of $\hbsx$ follows in a very similar manner. 
In order to compute $\hrsx$, consider picking a \textit{red} single edge uniformly at random (among all red single edges in $\mathbb{H}$) and assume that it is connected at one end to an {\em active} node.   %while the other end is {\em inactive}.
Then, we compute the generating function for the number of nodes influenced by following the other end of the edge, and obtain the following expression for the generating function $\hrsx$:
%Similar to \cite{YaganPRE}, we obtain the following expression for the generating function $\hrsx$:

\begin{align}
\hrsx &= x\sum_{\bd}\frac{\drs\pd}{<\drs>}\RrsFull D(\drs - 1, \nrt, \dbs, \nbt) \notag \\
        & ~~~~ + x^{0} \sum_{\bd}\frac{\drs\pd}{<\drs>} \lsb 1 -  \RrsFull \rsb
\label{eq:hrsx1}
\end{align}
where $D$ is as defined at (\ref{eq:defn_D}).

We now explain each term appearing at (\ref{eq:hrsx1}) in turn. The explicit factor $x$ stands for the initial vertex that is arrived at by following the randomly selected red single-edge. 
The term $\frac{\drs\pd}{<\drs>}$ gives the \textit{normalized} probability that the arrived vertex has colored degree $\bd$.
Since the arrived node is reached by a {\em red} link, it needs to be \textit{red}-vulnerable to be added to the vulnerable component.
If the arrived node is indeed \textit{red}-vulnerable, which happens with probability $\RrsFull$, it can activate other nodes via its remaining $\drs - 1$ \textit{red} single edges, $\dbs$ \textit{blue} single edges, $\nrt$ \textit{red} triangles, and $\nbt$ \textit{blue} triangles.
Because the number of vulnerable nodes reached by each of its \textit{red} single edges and triangles (resp.~\textit{blue} single edges and triangles) are generated in turn by $\hrsx$ and $\hrtx$ (resp. $\hbsx$ and $\hbtx$) respectively, we obtain the term $\hrsx^{\drs - 1}\hrtx^{\nrt}\hbsx^{\dbs}\hbtx^{\nbt}$ by the powers property of generating functions.
Averaging over all possible colored degrees $\bd$ gives the first term in (\ref{eq:hrsx1}).
The second term with the factor $x^{0}$ accounts for the possibility that the arrived node is {\em not} red-vulnerable and thus is not included in the cluster.
An analogous expression can be obtained for $\hbsx$ via similar arguments.

\subsection{Influence Propagation via {\em Red} Triangles}
\label{sec:influencePropagationTriangle}

We now derive $\hrtx$, i.e., the generating function for the number of nodes influenced by following a \textit{red} triangle selected at random; similar arguments hold for $\hbtx$.
%We demonstrate this situation in Figure (\ref{fig:triangleOne}) and (\ref{fig:triangleTwo}), where the top vertex $u$ is \textit{active}, and we are interested in computing the generating function for the number of nodes that will be influenced by nodes $v$ and $w$.
We demonstrate this situation in Figure \ref{fig:triangleInfluence}, where the top vertex $u$ is {\em active}, and we are interested in computing the generating function for the number of nodes that will be influenced by nodes $v$ and $w$.
We will compute the generating function $\hrtx$ by conditioning on the following events:
\begin{figure}[t]
	\centering
    \includegraphics[width=0.2\textwidth]{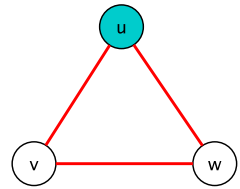}
    \caption{\sl
        Assume that the top vertex $u$ is {\em active} in network $\br$. Whether it will activate the other two nodes will depend on whether $v$ and/or $w$ are $\br$-vulnerable.
        }
	\label{fig:triangleInfluence}
\end{figure}

\begin{itemize}
\item
If neither of nodes $v$ and $w$ is \textit{$\br$}-vulnerable, then the number of nodes influenced will be zero.
Node $v$ has degree $\bd$
with \textit{normalized} probability $\frac{\nrt\pd}{<\nrt>}$, in which case it is not \textit{$\br$}-vulnerable with probability $1 -\RrsFull$. Similarly, 
the probability that node $w$ 
 has degree $\bd'$ and 
not $\br$-vulnerable is $(\frac{\nrt'\pdp}{<\nrt'>})(1 - \rho_{\bdp, rs})$.
Summing over all possible cases, we obtain the first term in (\ref{func:hrtx}) with $x^0$ (meaning that zero nodes will be influenced by following the red triangle in this case).

\item 
Consider the case where only one of $v$ and $w$ is influenced, leading to a term with the factor $x^1$ in (\ref{func:hrtx}).
Without loss of generality, consider the case where $v$ is activated while $w$ is not.
If node $v$ has degree $\bd$, then it is \textit{$\br$}-vulnerable with probability $\RrsFull$, and can influence other nodes in the usual manner.
Then, the event that node $w$, with degree $\bdp$, will not be activated despite having two active neighbors (nodes $u$ and $v$) has probability $1 - \rho_{\bdp, rt}$.
By symmetry and exchangeability of nodes $v$ and $w$, an equivalent term will be obtained for the case where $w$ is activated but $v$ is not. 
Summing over all possibilities we obtain the second term in (\ref{func:hrtx}).

\item Finally, we consider the case where both $v$ and $w$ become {\em active} giving rise to term with factor $x^2$.
        There are two possible scenarios:
    \begin{itemize}
        \item Both of $v$ and $w$ are $actived$ by $u$.
            The probability that $v$ is $actived$ by $u$ is
            $\RrsFull$
            as already discussed during the computation of the first term.
            By symmetry, the probability for $w$ is the same as for $v$.
            Multiplying the two probabilities leads to the third term in (\ref{func:hrtx}).

    \item Only one of  $v$ and $w$ is made \textit{active} immediately by $u$ while the other is not; e.g., say $v$ is activated but not $w$. 
        However, $w$ also gets activated by the joint influence from $u$ and $v$. With $\bd$ and $\bdp$ denoting the degree of $v$ and $w$, this happens with probability
        $(\RrsFull)(\RrtFullp - \RrsFullp)$.
       Here, the second term accounts for the fact that $w$ gets activated {\em only if} it has two (or, more)  active neighbors.
        Summing over all possibilities as before, and multiplying by two for the case where $v$ and $w$ are replaced, we  obtain the last term in (\ref{func:hrtx}).
        \end{itemize}
\end{itemize}

\subsection{Deriving the Condition for Global Cascades}
\label{sec:derivationConditionProbabilityGlobalCascades}
The discussion given in Section \ref{sec:influencePropagationSingle} and \ref{sec:influencePropagationTriangle} leads to a set of recursive  equations for $\hrsx$, $\hrtx$, $\hbsx$, and $\hbtx$. Recursions for $\hrsx$ and $\hrtx$ are given in (\ref{func:hrsx}) and (\ref{func:hrtx}), respectively; the expressions for $\hbsx$ and $\hbtx$ are very similar and omitted here for brevity.
With these four recursive equations in place, it is possible to determine the characteristic function $H(x)$ of the {\em finite} number of nodes activated in the contagion process. Namely, for a given $x$, we shall find a 
{\em fixed-point} of these recursive equations, and then use the resulting values of $\hrsx$, $\hrtx$, $\hbsx$, and $\hbtx$ in (\ref{eq:Hx}) to get $H(x)$.

\begin{widetext}
{\small
\begin{align}
    \hrsx &= x\sum_{\bd}\frac{\drs\pd}{<\drs>}\RrsFull D(\drs - 1, \nrt, \dbs, \nbt) + x^{0} \sum_{\bd} \frac{\drs\pd}{<\drs>}\lsb 1 -  \RrsFull \rsb
    \label{func:hrsx}
    \\
    \hrtx &= x^{0}\sum_{\bd} \sum_{\bdp}  \frac{\nrt\pd}{<\nrt>} \lsb 1 - \RrsFull \rsb \frac{\nrt'\pdp}{<\nrt'>} (1 - \RrsFullp)       \label{func:hrtx}
 \\
     & ~~~~ + 2x\sum_{\bd} \sum_{\bdp} \frac{\nrt\pd}{<\nrt>} \RrsFull D(\drs, \nrt - 1, \dbs, \nbt) \frac{\nrt'\pdp}{<\nrt'>} \lsb 1 - \RrtFullp \rsb \notag \\
     & ~~~~ + x^2 \sum_{\bd} \sum_{\bdp} \lsb \frac{\nrt\pd}{<\nrt>} \RrsFull D(\drs, \nrt - 1, \dbs, \nbt) \rsb \lsb \frac{\nrt'\pdp}{<\nrt'>} \RrsFullp D(\drs', \nrt' - 1, \dbs', \nbt') \rsb \notag \\
    & ~~~~ +2 x^2 \sum_{\bd} \sum_{\bdp} \lsb \frac{\nrt\pd}{<\nrt>} \RrsFull D(\drs, \nrt - 1, \dbs, \nbt) \rsb \lsb  \frac{\nrt'\pdp}{<\nrt'>} \lsb \RrtFullp - \RrsFullp \rsb D(\drs', \nrt' - 1, \dbs', \nbt') \rsb 
\notag
\end{align}
}
\end{widetext}

%With the solutions of the recursive equations (\ref{func:hrsx}) - (\ref{func:hbtx}) for any given $x$, i.e., a fixed point of them, we put them in (\ref{eq:Hx}) to get the probability of triggering a global cascade.

By conservation of probability and the definition of generating functions, we know that $H(1)=1$ only if final number of activated nodes is {\em finite} with probability one. In other words, {\em global} cascades that lead to a positive fraction of influenced nodes are possible only if $H(1) < 1$.  This prompts us to seek a fixed point of the recursive equations when $x = 1$.
For notational convenience, we define $h_1 := \hrsOne$, $h_2 := \hrtOne$, $h_3 := \hbsOne$, and $h_4 := \hbtOne$.
From (\ref{eq:Hx}), this gives
\begin{align}
    H(1) = \sum_{\bd}\pd h_1^{\drs}h_2^{\nrt}h_3^{\dbs}h_4^{\nbt},
    \label{eq:HxOne}
\end{align}
while the recursions take the form
\begin{align}
h_i = g_i(h_1, h_2, h_3, h_4), \quad i=1, 2, 3, 4.  
\label{eq:recursion_simplified}
\end{align}
Here the functions $g_1, g_2$ are easily obtained from (\ref{func:hrsx}) and (\ref{func:hrtx}), and similarly $g_3$ and $g_4$ can be obtained from the recursions for $\hbsx$, and $\hbtx$. To give an example, we have
\begin{align}
& g_3(h_1,h_2,h_3,h_4) 
\notag \\
 & = \sum_{\bd}\frac{\dbs\pd}{<\dbs>}\left(\RbsFull h_1^{\drs}h_2^{\nrt}h_3^{\dbs-1}h_4^{\nbt} +1 -  \RbsFull \right) 
 \notag
\end{align}

It is clear that the recursions (\ref{eq:recursion_simplified}) have a trivial fixed point $h_1 = h_2 = h_3 = h_4 = 1$ which yields $H(1) = 1$, meaning that  cascades will die out without reaching a positive fraction of the population with high probability. 
However, the trivial solution is {\em the physical solution} only if it is a stable fixed point.
To check its stability, we linearize the recursive equations  
(\ref{eq:recursion_simplified}) around $x=1$, and compute the corresponding 
 Jacobian matrix $\mathbf{J}$ via
 \begin{align}
     \mathbf{J}(i,j) = \frac{\partial g_i(h_1,h_2,h_3,h_4)}{\partial h_j} \Bigg|_{h_1=h_2=h_3=h_4=1}
 \end{align}
 for each $i,j =1,2,3,4$; the exact expression for the four by four matrix $\mathbf{J}$ is not give here in order to save space. 
 Now, the trivial solution $h_1=h_2=h_3=h_4=1$ is stable if and only if the largest eigenvalue in absolute value of $\mathbf{J}$, denoted $\sigma(\mathbf{J})$, is less than or equal to one. Otherwise, if $\sigma(\mathbf{J})>1$, then there exists another fixed point for the recursion with $h_1, h_2, h_3, h_4 < 1$, leading to $H(1) < 1$. In that case, the probability deficit $1-H(1)>0$ gives the probability that the contagion process reaches infinitely many nodes, i.e., a {\em global} spreading event takes place. Collecting, we conclude that the condition of global cascades is given by $\sigma(\mathbf{J})>1$, while the probability of global cascade equals $P_{trig}=1-H(1)$.

\section{Expected Cascade Size}
\label{sec:expectedCascadeSize}
Next, we are interested in computing the expected size of global cascades when they take place.
Put differently, we will analyze the expected fraction of nodes that will eventually become active as we randomly pick a node in the network and set it active.
We follow the approach used in \cite{gleeson2007seed, hackett2011cascades,YaganPRE}, which has been proven to be an effective way to analyze expected cascade size in networks.

First, consider the network $\mathbb{H}$ as a tree-structure with an arbitrary node selected as the root. 
Then, label the levels of the tree from $\ell=0$ at the bottom to $\ell = + \infty$ at the top of the tree. 
Similar to \cite{hackett2011cascades, YaganPRE}, we assume that nodes begin updating their states starting from the bottom of the tree and proceeding to the top.
In other words, we assume that a node at level $\ell$ updates its state only after all nodes at the lower levels $0, 1, \dots, \ell-1$ finish updating.
We define $\qrsl$ as the probability that a node at level $\ell$ of a tree, which is connected to its unique parent by a \textit{red} single edge, is active given that its parent at level $\ell+1$ is inactive.
Then, we consider a pair of nodes at level $\ell$ that together with their parent at level $\ell+1$ form a red triangle. Given that the parent is inactive, we let
 $\qrtlOne$  (resp.~$\qrtlTwo$) denote the probability that only one (resp.~both) of the two
 child nodes of this triangle is active.
 %;
%these are demonstrated as the second and third cases in Figure \ref{fig:differentTypeActiveEdge}, respectively. 
We define $\qbsl$, $\qbtlOne$, and $\qbtlTwo$ for {\em blue} edges in the same manner.
 
% node at level $\ell$, which is connected to its unique parent by a \textit{red} triangle edge, is {\em active}, given that the parent node is inactive. 
%For $\qrtlOne$, we assume that the other child of the parent node in the same triangles is is only one {\em active} nodes in the triangle, while both of the nodes at level $\ell$ are active for $\qrtlTwo$,
%Hence, the difference between $\qrtlOne$ and $\qrtlTwo$ is that whether two of the nodes in a triangle at level $\ell$ are both {\em active}.

According to our model, an {\em active} node is never deactivated, meaning  that $\qrsl$, $\qrtlOne$, $\qrtlTwo$, $\qbsl$, $\qbtlOne$, $\qbtlTwo$ are all non-decreasing.
Therefore, they will converge to $\qrslInf$, $\qrtlOneInf$, $\qrtlTwoInf$, $\qbslInf$, $\qbtlOneInf$, $\qbtlTwoInf$.
Then, the expected cascade size (i.e., the fraction of active individuals) $S$ is given by the probability that the arbitrary selected node at the top of the tree becomes {\em active}.
In order to computer $S$, we first derive recursive relations for $\qrsl$, $\qrtlTwo$, $\qrtlTwo$, $\qbsl$, $\qbtlTwo$, $\qbtlTwo$. We have

\begin{widetext}
{\small
    \begin{align}
        \qrsln &= \sum_{\bd} \frac{\drs\pd}{<\drs>} \sum_{i = 0}^{\drs-1}\sum_{j = 0}^{\nrt} \sum_{x = 0}^{j} \sum_{m = 0}^{\dbs} \sum_{n = 0}^{\nbt} \sum_{y = 0}^{n}  Q_{\ell} \lmb\lsb i, j, x, m, y, n \rsb, (\drs - 1, \nrt, \dbs, \nbt) \rmb   F\lmb (i, x, j - x, m,  y, n - y), \bd \rmb
        \notag \label{eq:qrsln}
        \\
        \qrtlnOne &= 2 \sum_{\bd, \bdp}\frac{\nrt\pd}{\nrt} \frac{\nrt'\pdp}{<\nrt'>} \sum_{i,i' = 0}^{\drs,\drsp}\sum_{j,j' = 0}^{\nrt - 1,\nrtp-1} \sum_{x,x' = 0}^{j,j'} \sum_{m,m' = 0}^{\dbs,\dbsp} \sum_{n, n' = 0}^{\nbt,\nbt'} \sum_{y, y' = 0}^{n,n'} 
        Q_{\ell} \lmb\lsb i, j, x, m, y, n \rsb, (\drs, \nrt - 1, \dbs, \nbt) \rmb
        \notag \\ 
  & ~~~~        
       Q_{\ell} \lmb\lsb i', j', x', m', y', n' \rsb, (\drsp, \nrtp - 1, \dbsp, \nbtp) \rmb  F\lmb (i, x, j - x, m,  y, n - y), \bd \rmb 
       \notag \\
       &~~~~ (1- F\lmb (i', x'+1, j' - x', m',  y', n' - y'), \bdp \rmb)
       \notag
 %       \end{align}
%        }
%\end{widetext}
%
%\begin{widetext}
%{\small
%        \begin{align}
\\
        \qrtlnTwo &= \sum_{\bd, \bdp}\frac{\nrt\pd}{\nrt} \frac{\nrt'\pdp}{<\nrt'>} \sum_{i,i' = 0}^{\drs,\drsp}\sum_{j,j' = 0}^{\nrt - 1,\nrtp-1} \sum_{x,x' = 0}^{j,j'} \sum_{m,m' = 0}^{\dbs,\dbsp} \sum_{n, n' = 0}^{\nbt,\nbt'}\sum_{y, y' = 0}^{n,n'} 
        Q_{\ell} \lmb\lsb i, j, x, m, y, n \rsb, (\drs, \nrt - 1, \dbs, \nbt) \rmb
        \notag \\ 
  & ~~~        
        Q_{\ell} \lmb\lsb i', j', x', m', y', n' \rsb, (\drsp, \nrtp - 1, \dbsp, \nbtp) \rmb  \bigg(F\lmb (i, x, j - x, m,  y, n - y), \bd \rmb  F\lmb (i', x'+1, j' - x', m',  y', n' - y'), \bdp \rmb
       \notag \\
       & ~~~~ 
       + \big(F\lmb (i, x+1, j - x, m,  y, n - y), \bd \rmb - F\lmb (i, x, j - x, m,  y, n - y), \bd \rmb \big)  F\lmb (i', x', j' - x', m',  y', n' - y'), \bdp \rmb \bigg)
       \notag
        %\\
        %\qrsln &= \sum_{\bd}\frac{\drs\pd}{\drs} \sum_{i = 0}^{\drs}\sum_{j = 0}^{\nrt} \sum_{m = 0}^{\dbs - 1}\sum_{n = 0}^{\nbt} \bf\lmb\lsb i, j, m, n \rsb, (\drs, \nrt, \dbs - 1, \nbt), \tau, \ell \rmb\\
        %\qbtlnOne &= \sum_{\bd}\frac{\nbt\pd}{\nbt} \sum_{i = 0}^{\drs}\sum_{j = 0}^{\nrt} \sum_{m = 0}^{\dbs}\sum_{n = 0}^{\nbt - 1} \bf\lmb\lsb i, j, m, n \rsb, (\drs, \nrt, \dbs, \nbt - 1), \tau, \ell\rmb\\
        %\qbtlnTwo &= \sum_{\bd}\frac{\nbt\pd}{\nbt} \sum_{i = 0}^{\drs}\sum_{j = 0}^{\nrt} \sum_{m = 0}^{\dbs}\sum_{n = 0}^{\nbt - 1} \bf\lmb\lsb i, j, m, n \rsb, (\drs, \nrt, \dbs, \nbt - 1), \tau, \ell \rmb
        %\label{eq:qbtln}
    \end{align}
}    
where we define 
{\small
\begin{align}
     Q_{\ell} \lmb\lsb i, j, x, m, n, y  \rsb, (d_{1}, d_{2}, d_{3}, d_{4}) \rmb &= {d_{1}\choose i} \qrsl^{i}(1 - \qrsl)^{d_{1} - i}  {d_{2}\choose j}  {j \choose x} \qrtlOne^{x} \qrtlTwo^{j-x} (1 - \qrtlOne - \qrtlTwo)^{d_{2} - j}  {d_{3}\choose m} \qbsl^{m}
        \notag \\
    & ~~~~~~~ (1 - \qbsl)^{d_{3} - m} {d_{4}\choose n}  {n \choose y}  \qbtlOne^{y} \qbtlTwo^{n - y} (1 - \qbtlOne - \qbtlTwo)^{d_{4} - n}  
\end{align}
}
\end{widetext}

In words, $Q_{\ell} \lmb\lsb i, j, x, m, n, y  \rsb, (d_{1}, d_{2}, d_{3}, d_{4}) \rmb$ gives the probability that a node at level $\ell$ with colored degree $(d_1,2d_2,d_3,2d_4)$ has
\begin{itemize}
    \item $i$ (resp.~$d_1-i$) of the $d_1$ neighbors connected through red single edges as active (resp.~inactive). Similarly, $m$ (resp.~$d_3-m$) of the $d_3$ neighbors connected through blue single edges as active (resp.~inactive)
     \item of the $d_2$ red-triangles it participates in, $x$ has one active node, $j-x$ has two active nodes, and $d_2-j$ has no active node. Similarly,
     of the $d_4$ blue-triangles it participates in, $y$ has one active node, $n-y$ has two active nodes, and $d_4-n$ has no active node
\end{itemize}
Hence, multiplying 
$Q_{\ell} \lmb\lsb i, j, x, m, n, y  \rsb, (d_{1}, d_{2}, d_{3}, d_{4}) \rmb$ with $F\lmb (i, x, j - x, m,  y, n - y), \bd \rmb$ 
and summing over all possibilities for $\bd$ and $i,j,x,m,n,y$
gives the probability that the node under consideration turns active. This confirms the first expression above. Second and third terms consider simultaneously a pair of nodes that are part of a red triangle (where the top, i.e., parent, vertex is inactive). Therefore, we first condition on 
the degrees of these two nodes being $\bd$ and $\bdp$ respectively, and consider all possibilities concerning the states (active vs.~inactive) of these neighbors. Then for $\qrtlnOne$, we realize by symmetry that the desired expression is two times the probability that the node with degree $\bd$ turns active, and despite having one extra active neighbor, the node with degree $\bdp$ does not turn active. The fact that first node turns active is incorporated in the expression 
$(1- F\lmb (i', x'+1, j' - x', m',  y', n' - y'), \bdp \rmb)$
by the term $x'+1$.
For $\qrtlnTwo$, we proceed similarly and realize that for both nodes to turn active there are two possibilities. The node with degree $\bd$ either turns active regardless of the state of the node with degree $\bdp$ (in which case the node with degree $\bdp$ will turn active with probability $ F\lmb (i', x'+1, j' - x', m',  y', n' - y'), \bdp \rmb$), or it turns active only after the node with degree $\bdp$ does.

With the above recursion in place, we compute the final cascade size via
\begin{widetext}
{\small
\begin{align}
    S = \sum_{\bd}\pd  \sum_{i = 0}^{\drs}\sum_{j = 0}^{\nrt} \sum_{x = 0}^{j} \sum_{m = 0}^{\dbs} \sum_{n = 0}^{\nbt} \sum_{y = 0}^{n}   Q_{\infty} \lmb\lsb i, j, x, m, n, y \rsb, (\drs, \nrt, \dbs, \nbt) \rmb F\lmb (i, x, j - x, m,  y, n - y), \bd \rmb
    \label{eq:sSize}
\end{align}
}
\end{widetext}
%To check the stability, we can also use the same way as Section %\ref{sec:derivationConditionProbabilityGlobalCascades}.
%That is, we firstly get the Jacobian matrix obtained from (\ref{eq:qrsln}), then check whether the largest absolute value of equal 1.
Namely, we first solve for the values of $\qrslInf$, $\qrtlOneInf$, $\qrtlTwoInf$, $\qbslInf$, $\qbtlOneInf$, $\qbtlTwoInf$ using the recursive equations, and then substitute them into (\ref{eq:sSize}) to obtain the expected size of global cascades.

%%Under the natural condition F((0, 0), k) = 0, q1,‚àö¬¢¬¨√†¬¨√ª =
%%q2,‚àö¬¢¬¨√†¬¨√ª = 0 is the trivial fixed point of the recursive equations
%(11)-(12). In view of (13), this trivial solution yields
%S = 0 pointing out the non-existence of global spreading
%events. However, the trivial fixed point may not be stable
%and another solution with q1,‚àö¬¢¬¨√†¬¨√ª, q2,‚àö¬¢¬¨√†¬¨√ª > 0 may exist.
%%In fact, the condition for the existence of a non-trivial
%solution can be obtained by checking the stability of the
%%trivial fixed point via linearization at q1,` = q2,` = 0. The

%\subsection{Relationship to Prior Work}

\section{Numerical Results}
\label{sec:numericalResults}
\subsection{Networks with Doubly Poisson Distributions}
\label{sec:networksWithDoublyPoissonDistributions}
%\begin{figure*}[ht]
%    \centering
%    \subfigure[]{\hspace{-0.5cm} \includegraphics[totalheight=0.2\textheight,width=0.43\textwidth] {figs/cascade_ptrigger_c025.eps}}
%    \subfigure[]{\hspace{2cm} \includegraphics[totalheight=0.2\textheight,width=0.43\textwidth] {figs/cascade_ptrigger_c1.eps}}
%    \label{fig:expTriggerCascadeSizeC1}
%    \caption{
%            Simulation for doubly Poisson degree distributions.
%            We set the content parameter and the threshold as \cite{YaganPRE}, $c = 0.25$ and $\tau^{*} = 0.18$. 
%    }
%\label{fig:expTriggerCascadeSize}
%\end{figure*}

In our first simulation study,
we use doubly Poisson distribution for the number of \textit{single edges} and \textit{triangles} in both networks.
%; i.e., $p_{st}^{r}$ and $p_{st}^{b}$ are independent Poisson distribution and independent.
Namely, we set
\begin{align}\nonumber
p_{st}^{r} &= e^{-\lambda_{r,1}}\frac{(\lambda_{r,1})^{s}}{s!}e^{-\lambda_{r,2}}\frac{(\lambda_{r,2})^{t}}{t!}, \mbox{ \ s, t = 0, 1, \dots },
%\label{func:poiPrst}
\\ \nonumber
p_{st}^{b} &= e^{-\lambda_{b,1}}\frac{(\lambda_{b,1})^{s}}{s!}e^{-\lambda_{b,2}}\frac{(\lambda_{b,2})^{t}}{t!}, \mbox{ \ s, t = 0, 1, \dots },
%\label{func:poiPbst}
\end{align}
where $s$ and $t$ are the number of \textit{single edges} and \textit{triangles} in the corresponding networks, respectively. Thus,  $\lambda_{r,1}$ and $\lambda_{r,2}$ (resp.~$\lambda_{b,1}$ and $\lambda_{b,2}$) denote the mean number of single edges and triangles, respectively in $\br$ (resp.~in $\bb$).

\begin{figure}[!t]
    \centering
    \subfigure[]{\hspace{-0.5cm} \includegraphics[totalheight=0.18\textheight,width=0.44\textwidth] {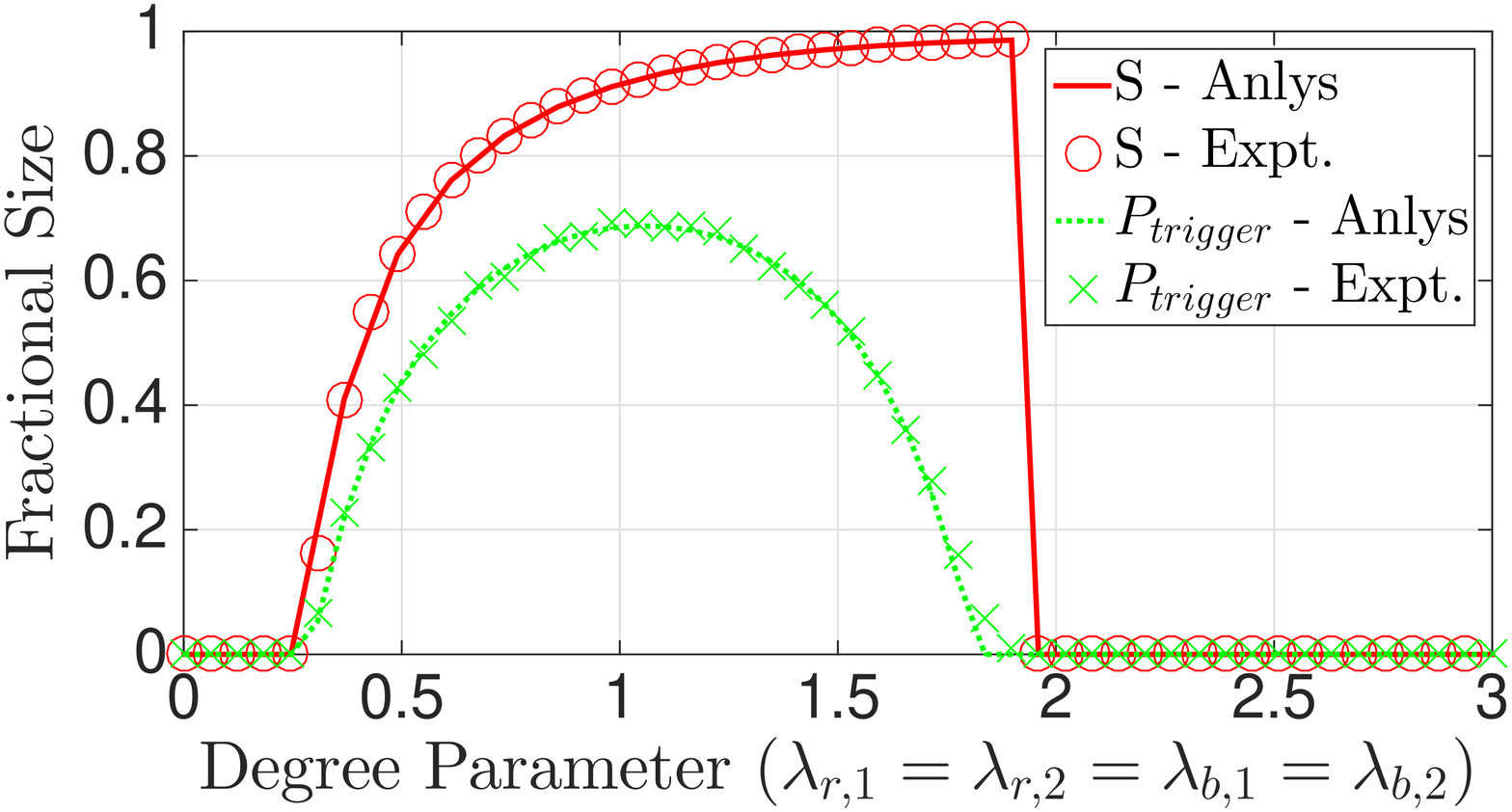}}
    \label{fig:expTriggerCascadeSizeC025_o}
    \subfigure[]{\hspace{-0.5cm} \includegraphics[totalheight=0.18\textheight,width=0.44\textwidth] {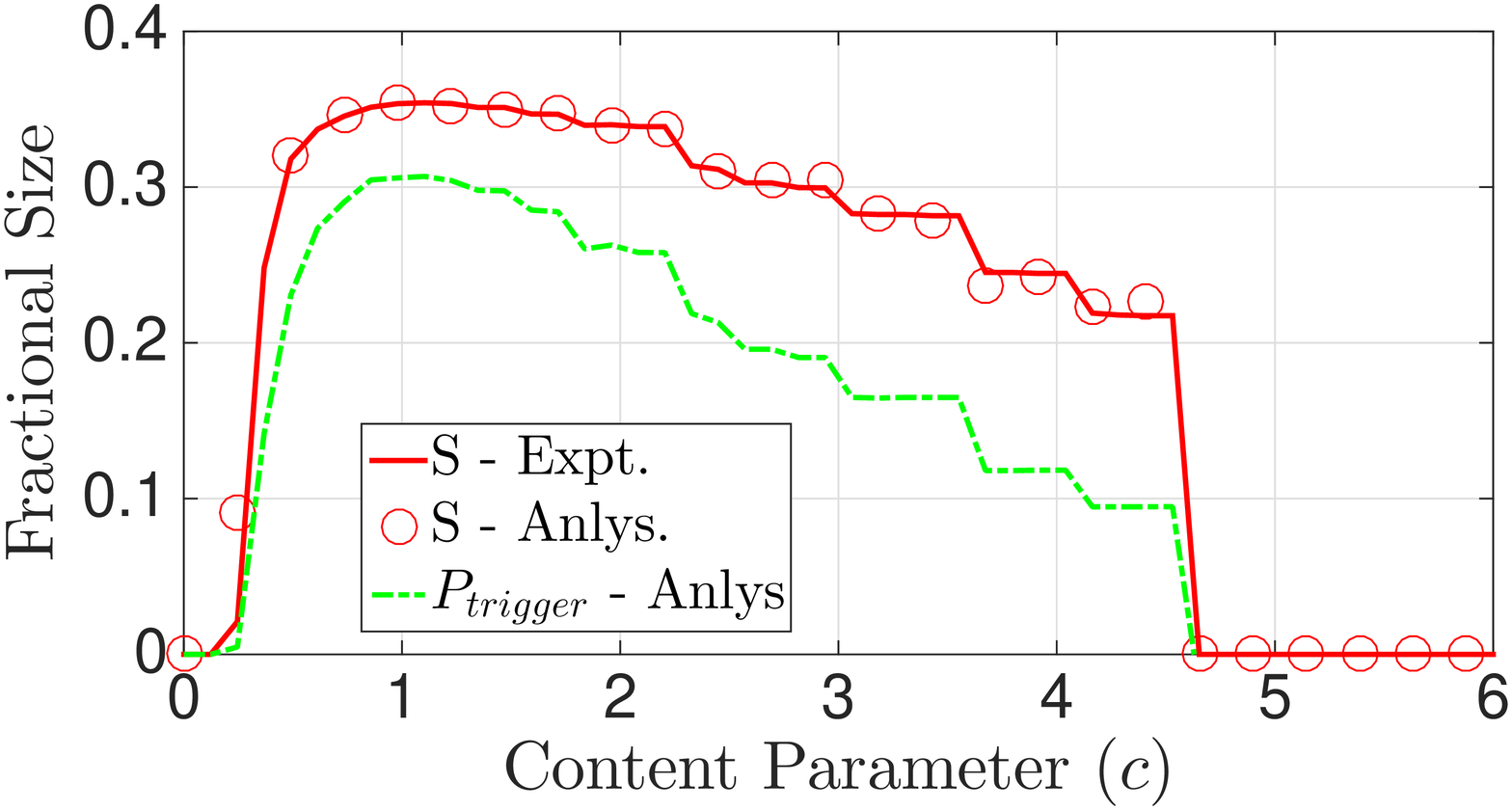}}
    \label{fig:exp_content_parameter}\vspace{-2mm}
    \caption{\sl
            Simulations for doubly Poisson degree distributions.
            In (a), we set the content parameter $c = 0.25$, the threshold as  $\tau = 0.18$, and $\alpha = 0.5$, and vary the degree parameters.
            In (b), we fix $\tau = 0.18$, $\lambda_{r, 1} = \lambda_{r, 2} = \lambda_{b, 1} = \lambda_{b, 2} = 0.3$, and $\alpha = 0.5$ while varying content parameter $c$.
    }
\label{fig:expTriggerCascadeSize}
\end{figure}

We consider $n = 1 \times 10^{5}$ nodes in the population and $\alpha = 0.5$ for the size of network $\br$.
We let $\tau = 0.18$ and $c = 0.25$ for the threshold and content parameters, respectively. 
The results are shown in Figure \ref{fig:expTriggerCascadeSize} where the curves stand for the theoretical results of probability $P_{trig}$ and expected size $S$ of cascades (obtained from our discussion in Section \ref{sec:conditionProbabilityGlobalCascades} and \ref{sec:expectedCascadeSize}), as a function of $\lambda_{r,1}=\lambda_{r,2}=\lambda_{b,1}=\lambda_{b,2}$. 
The markers stand for the empirical results for the same quantities, and are obtained by averaging over  5,000 independent experiments.
We see  a very good agreement between the analytical and experimental results confirming the validity of our analysis.
The slight discrepancy observed in $P_{trig}$ is due to the limited number of experiments, and can be mitigated by increasing the number of realizations. 
%\footnote{For $P_{trig}$, because of the difficulty of calculating the size of the extended component EGIN, we use 5,000 realizations to obtain $P_{trig}$.
%Namely, we create a multiplex network $\bh$, randomly pick up a node, assign a content to it, and check whether there is a global cascade.
%Then, we let the fraction of the case where there is a cascade as our $P_{trig}$.
%That's why there is some tiny fluctuations between the analytical $P_{trig}$ and the experimental.}

Next, we change our experimental set-up to demonstrate the effect of content parameter on the probability and size of cascades.
To that end, we fix all network parameters 
and observe the quantities of interest as the content parameter $c$ varies.
In particular, we set $\lambda_{r,1}=\lambda_{r,2}=\lambda_{b,1}=\lambda_{b,2}= 0.3$ and $\tau = 0.18$. We see that the probability and expected size of global cascades vary greatly as $c$ changes. This can be taken as an indication that our model can capture the real-world phenomenon that over the same population certain contents can become widespread while others die out quickly. 
In the setting used here, we see that global cascades take place when the content parameter $c$ is not too small or large.
The reason is that with a too small or large content parameter, the connectivity of the conjoined network is dominated by only one of the two networks.
So, if neither of them has enough connectivity to trigger a global cascade by their own, then there will be no global cascades in the conjoined network.  
%for a small or large content parameter $c$.
When the $c$ is neither too large nor too small (e.g., close to unity), both networks will contribute to the connectivity together and it becomes possible to trigger a global cascade.
For other values of $\lambda_{r}$ and $\lambda_{b}$ a completely different situation might occur, e.g., with very small or very large $c$ promoting cascades; e.g., see \cite[Fig. 2]{YaganPRE} for a few such examples.

\begin{figure*}[!t]
    \centering
    \subfigure[]{\hspace{-0.3cm} \includegraphics[width=0.32\textwidth] {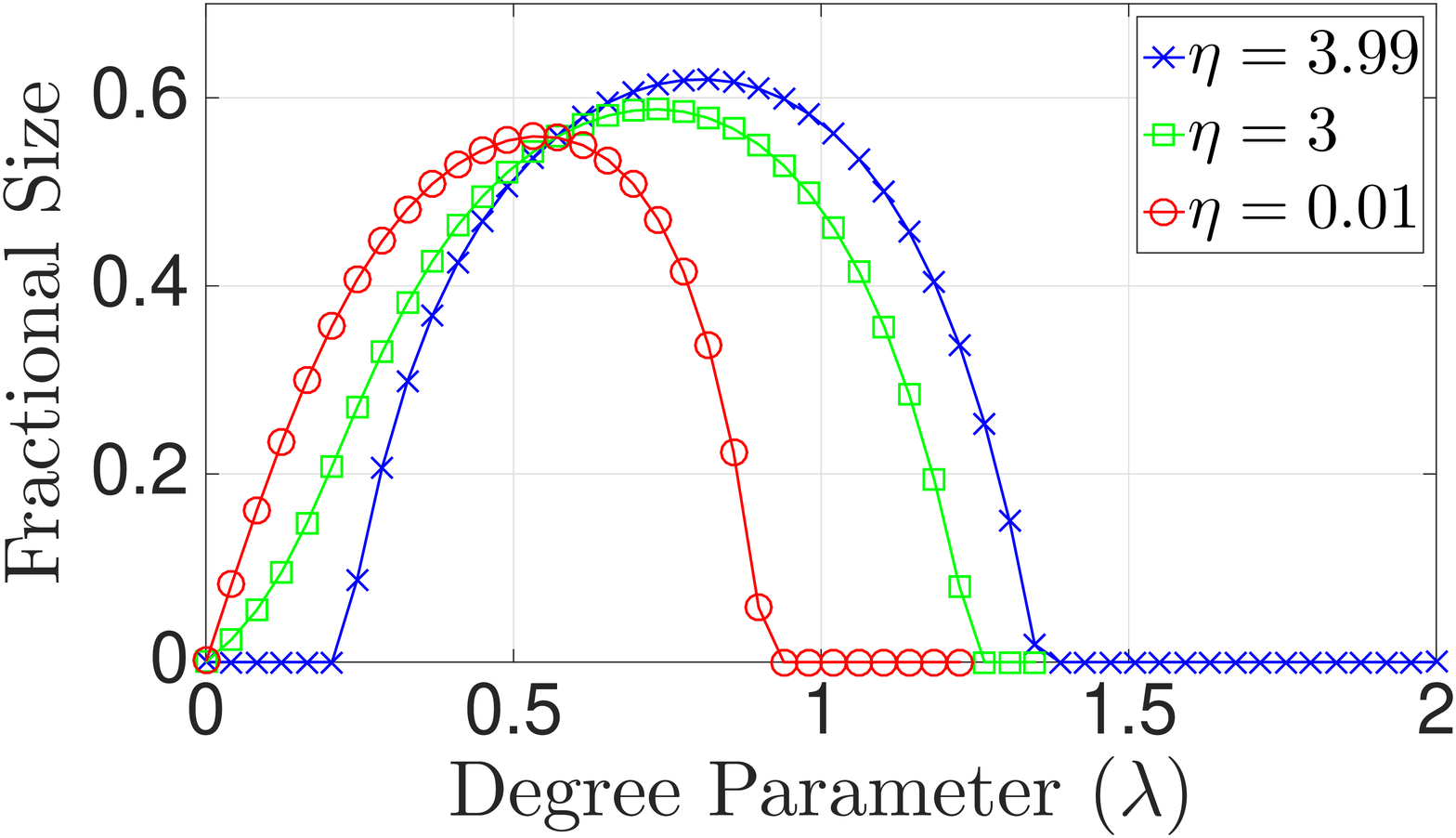}}
    \label{fig:expTriggerCascadeSizeC025}
    \subfigure[]{\hspace{0.1cm} \includegraphics[width=0.32\textwidth] {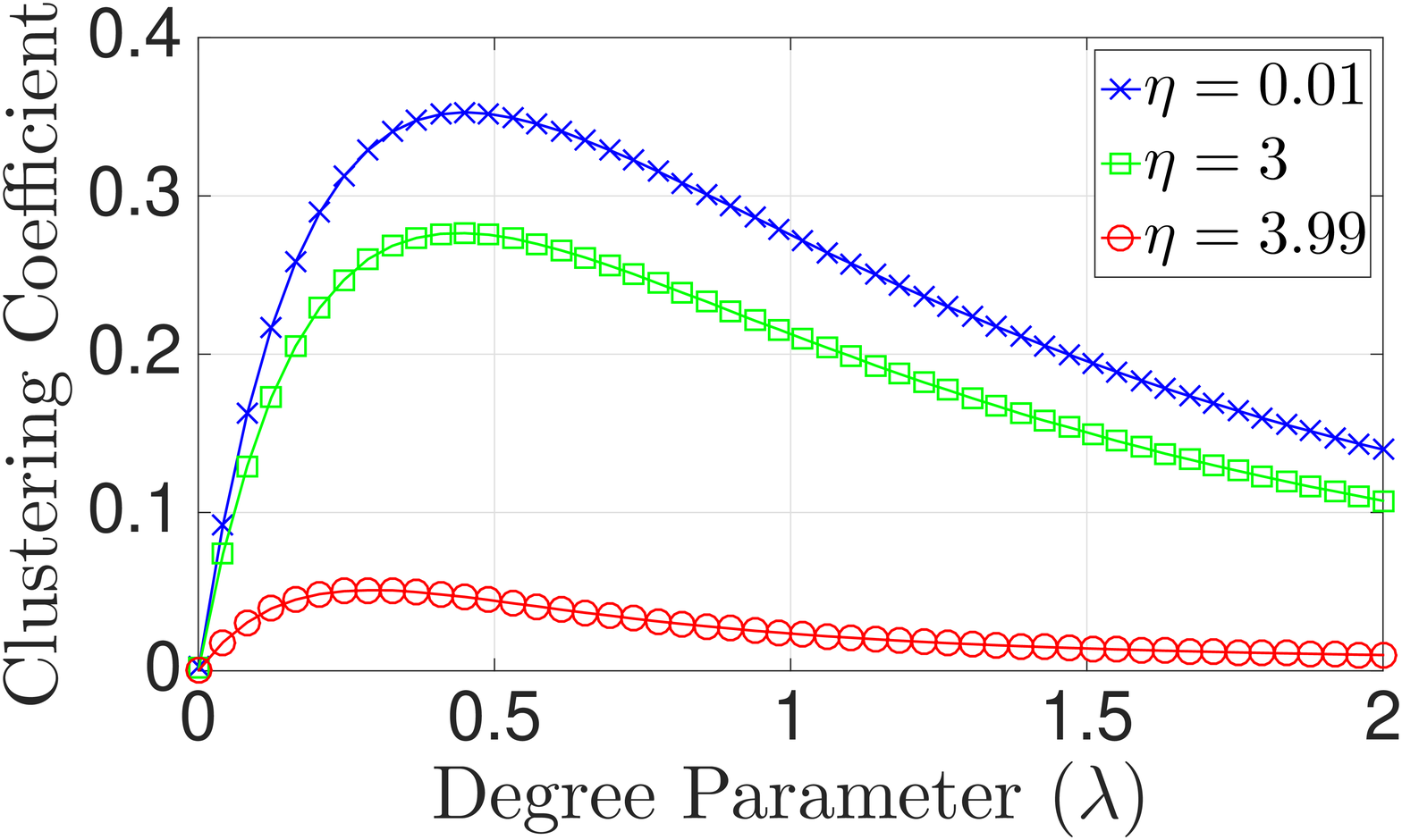}}
    \label{fig:expTriggerCascadeSizeC1}
    \subfigure[]{\hspace{0.1cm} \includegraphics[width=0.32\textwidth] {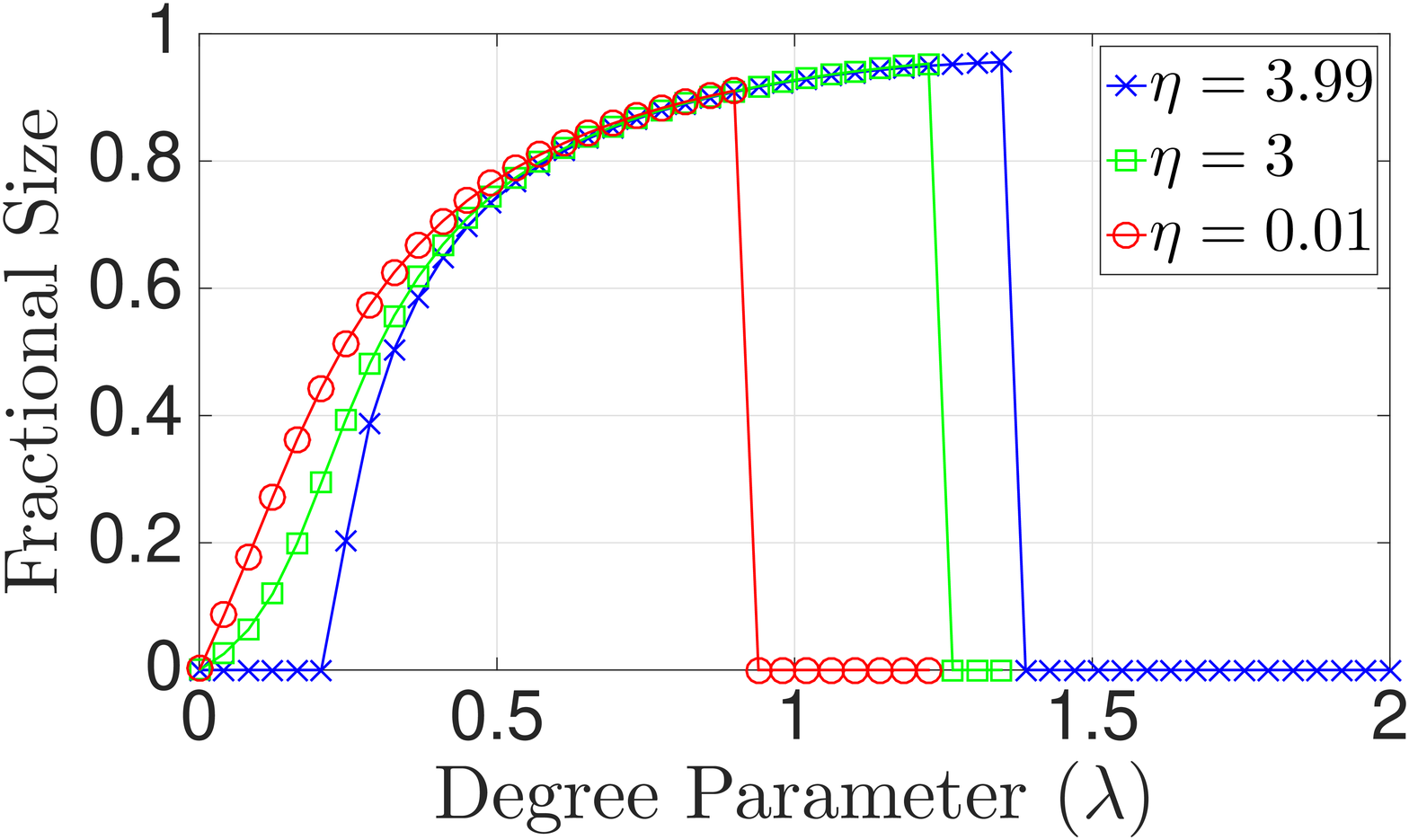}}
    \label{fig:expTriggerCascadeSizeC1_o}
    \caption{\sl Illustration of the effect of clustering coefficient on the expected cascade and probability of global cascades.
    We fix $\tau = 0.18$, $c = 0.25$, and $\alpha = 0.5$, then vary the degree parameter $\lambda$ defined in Table \ref{table:parametersDoublyPoissionDistribution}. We see
  (a) the probability to trigger a global cascade;
  (b)  the global clustering coefficient introduced in Section \ref{sec:networkClustering}; and
  (c) the expected cascade size.
  %What is $\tau$, what is $c$??
    }
\label{fig:expTriggerCascadeSize2}
\end{figure*}

\subsection{How does Clustering Affect the Cascade Size?}
Our next goal is to reveal the impact of clustering on the influence propagation process.
In order to control the level of clustering while keeping the mean total degree fixed,
we use Poisson distributions for the number of single and triangle edges in two networks with parameters
given in Table \ref{table:parametersDoublyPoissionDistribution}. 
Obviously, the clustering coefficient in $\br$ is fixed while with $\eta \in [0, 4]$ the clustering of $\bb$ varies between the two extremes: i) when $\eta = 4$, $\bb$ will have no {\em single}-edges and consist only of \textit{triangles} resulting with a clustering coefficient close to one; and ii)
with $\eta = 0$, there will be no triangles in $\bb$ and hence its clustering coefficient will be close to zero.
Collecting, we see that the clustering coefficient of $\bb$ and thus of $\bh$
increases with increasing $\eta$ in this setting.

\begin{table}[!h]
    \centering
    \begin{tabular}{l|c|c}
                                            &   Network $\br$   &   Network $\bb$\\
        \hline
        Distribution of \textit{single}-edges   &   Poi($2\lambda$)  &   2  Poi$( \frac{4- \eta}{2} \lambda)$              \\
        & & \\
       Distribution of \textit{triangles} &   Poi($\lambda$)   &   Poi$\left(\frac{\eta}{2} \lambda \right)$
    \end{tabular}
    \caption{\sl Parameters of the doubly Poisson distribution. This choice ensures that the mean and variance of the total degree distribution (single plus triangle edges) in $\bb$ are independent of $\eta$, while its clustering varies greatly as $\eta$ varies in $(0,4)$. In Figure \ref{fig:expTriggerCascadeSize} we set $\lambda = 0.5$.
%With the help of it, we can focus on the impacts of clustering without involving more factors.
    %We use $\lambda_{\bf} = 0.36$ and $\lambda_{\bw} = 0.5$ for Figure \ref{fig:expCascadeRegion}.
    }
    \label{table:parametersDoublyPoissionDistribution}
\end{table}

With these in mind, we first demonstrate the impact of clustering on the probability of triggering a global cascade.
Figure \ref{fig:expTriggerCascadeSize2}(a) shows the probability of triggering a global cascade as a function of $\lambda$ for three different $\eta$ values. The resulting clustering coefficients 
are plotted in Figure \ref{fig:expTriggerCascadeSize2}(b) where we clearly see that clustering increases with 
increasing $\eta$. The main observation from 
Figure \ref{fig:expTriggerCascadeSize2}(a) is that 
increasing the clustering (i.e., increasing $\eta$), shifts the interval of $\lambda$ for which global cascades are possible to the {\em right}. This leads to a double-faceted conclusion that 
clustering decreases the probability of global cascades when average degrees are small, whereas after a certain value of average degree, clustering increases the probability of cascades. 

The double-faceted impact of clustering on cascade probability can be explained as follows. It is known \cite{WattsExternal, OY13} that threshold models of complex contagion exhibit two phase transitions as the average degree increases, a second-order transition at {\em low} degrees that marks the formation of a giant vulnerable component and a first-order transition at {\em high} degrees due to increased {\em local} stability of nodes; namely, due to the increased difficulty of activating high-degree nodes. 
Given that clustering is known to decrease the size of giant component \cite{zhuang2015information}, 
we expect that it will be more difficult for a clustered network to contain a giant vulnerable cluster. This is why the lower phase transition in complex contagions appear later (i.e., at larger degrees) as clustering increases. On the other hand, the cycles of size three (i.e., triangles) that are common in clustered networks
can help trigger cascades when average degree is higher. For instance, in a tree-like network a single active node can only activate its vulnerable connections. However, in a triangle, an active node may first activate one of its vulnerable connections, making it possible for the third node to be activated (which now has two active neighbors) even if it is {\em not} vulnerable. This is what pushes the second phase transition to higher degrees.

%If an \textit{active} node active one of its neighbors in the triangle, the third one will have a higher probability to be activated as two of its neighbors are \textit{active}.
%However, in a non-clustered network, if an \textit{active} make one of its neighbors \textit{active}, then the two \textit{active} nodes cannot try to active the third nodes simutaneously as they do in clustered networks because there is no cycles of size three.
%With this in mind, it becomes obvious that why there still a global cascade for clustered networks when the average degree is large.

Next, we explore the impact of clustering on the expected cascade size in Figure \ref{fig:expTriggerCascadeSize}(c).
Here again, we see 
the double-faceted impact of clustering with small average degrees favoring low clustering, while high degrees favoring high clustering in terms of having a larger cascade size. In fact, we see the existence of a critical average degree (around $\lambda=0.6$ in Figure \ref{fig:expTriggerCascadeSize}(c)) such that when $\lambda$ is smaller (resp.~larger) than the critical value, expected cascade size decreases (resp.~increases) with increasing clustering.

%In addition, under the same average degree, the expected size decreese with the increasing of clustering if there exits a global cascade i.e., $P_{trig}$ is not zero.
%This effect of clustering on the diffusion process is attributed to the fact that there may be redundant edges in triangles.
%For example, if one of nodes in a triangle can active both of its neighbors in the same triangle, then the edges betwee the two actived nodes is redundant.

%If we keep increasing the average degree a very large number, then the global cascade in all of them will disappear due to the high connectivity.

%\begin{figure}[!ht]
%	\centering
%    \includegraphics[width=0.45\textwidth]{figs/cascade_clustering.eps}
%    \caption{
%        Comparison of the probability and expected size of a global cascade between different level of clustering.
%        Plots are obtained from our analytical results as we have shown the agreement between the analytical and the experimental in Section 
        %\ref{sec:networksWithDoublyPoissonDistributions}.
%        We use the same content parameter and threshold as the experiments in Fig. 
%        \ref{fig:expTriggerCascadeSizeC025}
%        $c = 0.25$, $\tau^{*} = 0.18$ and $\alpha = 0.5$.
%        }
%	\label{fig:expImpactClustering}
%\end{figure}

\begin{figure}[!h]
	\centering
    \includegraphics[totalheight=0.20\textheight, width=0.48\textwidth]{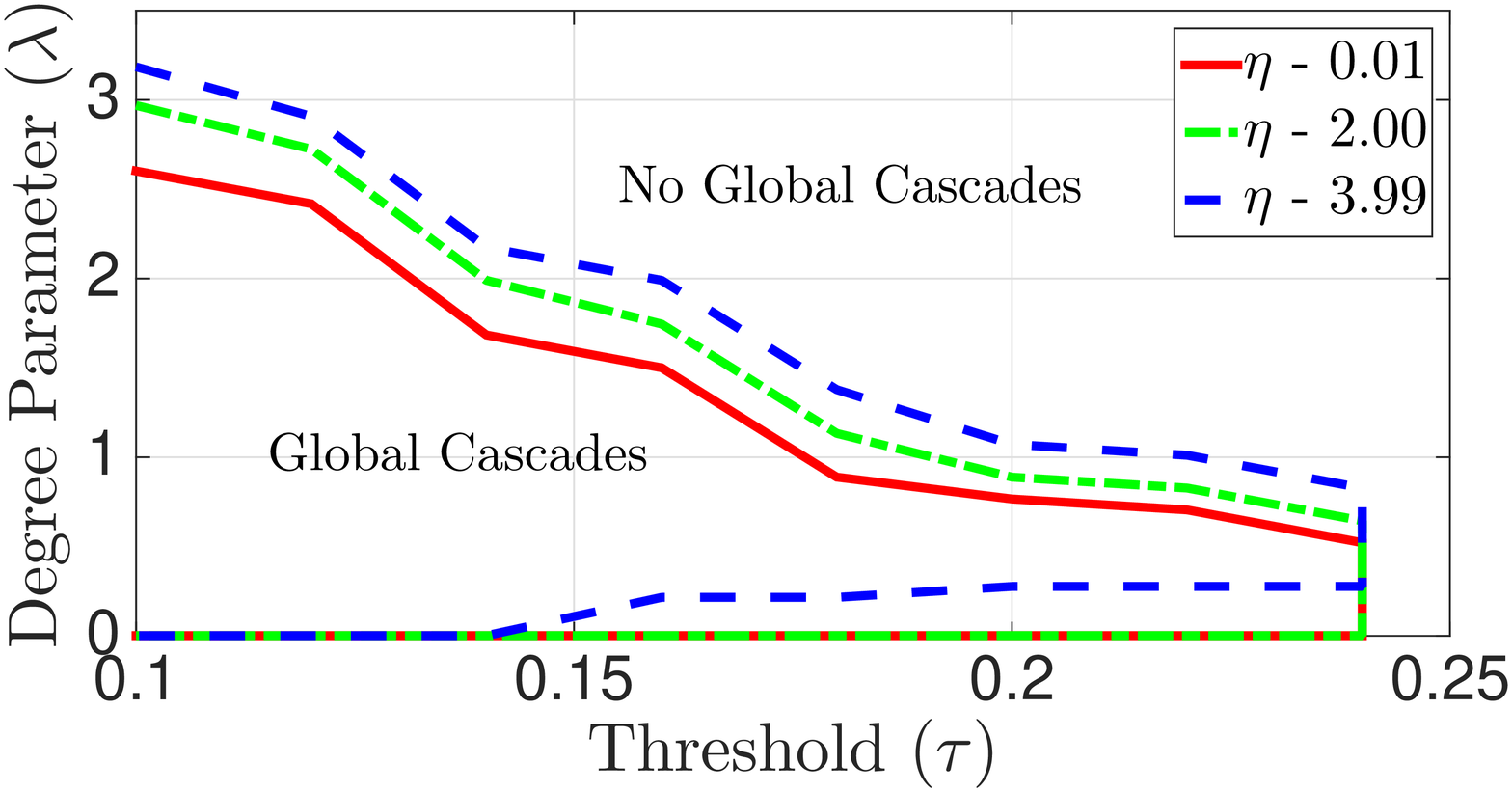}
    \caption{\sl
        We show the {\em cascade regions} in the Degree Parameter-Threshold plane when $\alpha = 0.5$, $\tau = 0.18$,
        and both networks follow doubly Poisson distributions as described in Table \ref{table:parametersDoublyPoissionDistribution}. Clustering increases as $\eta$ increases. %from
       %$0.01$ to $2$ to $3.99$.
       %
       %
       % 
        %Because the planes separate the region between whether there exists a global cascade.
        %In other words, the regions that fall inside the lines correspond to the case where $\sigma(J) > 1$, while we have $\sigma(J) \leq 1$ outside the boundary.
        }
	\label{fig:expCascadeRegion}
\end{figure}

Finally, we consider the impact of clustering on the average degree-cascade threshold plane.
For each parameter pair $(\lambda, \tau)$, the curves in Figure \ref{fig:expCascadeRegion} separate the region where global cascade can take place (areas inside the boundaries) from the region where they cannot (areas outside the boundaries).
Once again, we confirm that increasing the clustering coefficient shifts the interval where cascades are possible {\em up} (i.e., to higher degrees) for any threshold $\tau$. 

\section{Comparison between monoplex and Multiplex Networks}
\label{sec:comp_monoplex_multiplex}
In what follows, we will compare the dynamics of complex contagions
 over a monoplex network with that over a multiplex network. Of particular interest will be to find out whether the projection of
a multiplex network into a monoplex network leads to any significant differences in the dynamics that would warrant the separate analyses of multiplex networks as conducted here.

To identify the factors affecting complex contagions, we consider two 
different degree distributions to generate the networks.
In Section \ref{sec:networkWithLimitedAssortativity}, we use a setting similar to previous sections, with the resulting networks having almost no degree-degree correlations; e.g., assortativity defined as the Pearson correlation coefficient between the degrees of pairs of linked nodes \cite{newman2002assortative}.
In Section \ref{sec:networkWithAssortativity} and \ref{sec:networkMultiplePhases}, we use a different setting that leads to (tunable) assortativity for multiplex networks.
%In all cases, we will compare complex contagions in monoplex networks with that in multiplex networks.
In order to keep the focus on the comparison between monoplex and multiplex networks, we shall consider only non-clustered networks in the following discussion.

\subsection{Multiplex Networks with Limited Assortativity}
\label{sec:networkWithLimitedAssortativity}

First we consider the limited assortativity case and
%discuss the difference between monoplex and multiplex networks with limited assortativity.
 use the following degree distribution to assign blue and red stubs to each node: %(see also \cite{zhuang2015information}):
\begin{align}
\label{eq:degree_dist_colored_osy}
    p_{k}^{b} &= e^{-\lambda_{b}}\frac{\lambda_{b}^{k}}{k!}, \quad \mbox{ \ k = 0, \dots },    
    \\ \nonumber
    p_{k}^{r} &= \alpha e^{-\lambda_{r}}\frac{\lambda_{r}^{k}}{k!} + (1 - \alpha)\mathbf{1}[k = 0], \quad \mbox{ \ k = 0, \dots }.
\end{align}
A multiplex network is generated using the {\em colored} configuration model \cite{soderberg2003random,soderberg2002general} where stubs that are of the {\em same} color are matched randomly. The monoplex projected theory {\em ignores} the color of the edges and matches all stubs randomly with each other. An important question is whether we lose any significant information about contagion dynamics 
when the monoplex projected theory is used instead of the multiplex theory developed here and in \cite{YaganPRE}. For convenience and fair comparison, we set $\lambda_b=\lambda_r$ and use $c=1$ as the content parameter.

%we project them onto  the theory from monoplex networks.

%For multiplex networks, we use the same approach as in the previous sections.
%Clearly, the average degree of the monoplex network is the same as that of the multiplex network.
%The only difference between these two types of networks is the way to connect the stubs.

In Figure \ref{fig:expmonoplexVSMultiplexLimitedAssortativity01}, we set
$\alpha = 0.99$.
We see nearly no difference between the theoretical cascade sizes obtained from monoplex and multiplex theories, and they both match the simulation results well.
However, when $\alpha$ is reduced to 0.1 in Figure \ref{fig:expmonoplexVSMultiplexLimitedAssortativity099}, we clearly see a difference between the two theories and only multiplex theory matches the simulation results. This shows that even in the simplest case where both link types have the same influence factor (i.e., $c=1$),
monoplex theory may be unable to capture certain properties of cascade dynamics, reinforcing the need for studying cascades using the multiplex theory.

We now explain why the two cases, $\alpha = 0.1$ and $\alpha = 0.99$, lead to different conclusions
regarding the accuracy of the monoplex theory in capturing contagion dynamics over multiplex networks.
One of the key differences between the two cases is the resulting assortativity. 
In the former case with $\alpha = 0.1$, only $10\%$ of the nodes has red stubs, each of which shall be connected with other red stubs in the multiplex network case. Put differently, in this setting a small fraction of the population will have statistically {\em higher} degrees than the rest, and the additional links they have can only connect nodes with high degrees together. This leads to a positive correlation (i.e., assortativity) between the degrees of pairs of connected nodes. However, in the monoplex projection, the additional edges can be used to connect any two nodes, resulting with very little to no assortativity in the network. Obviously, when $\alpha$ is close to one, almost every node will have the additional edges and the above phenomenon will not be observed. Our simulation results confirm this intuition as we see that assortativity is negligible ($\sim 10^{-4}$) in both monoplex and multiplex cases when $\alpha=0.99$, while with $\alpha=0.1$, assortativity varies (as $\lambda_r=\lambda_b$ increases) from $0.05$ to $0.2$ in the multiplex case while still being negligible in the monoplex case. 

The impact of assortativity on the comparison between monoplex and multiplex theories is investigated further in the forthcoming discussion.

%As we can observe from Table \ref{table:statsmonoplexMultiplexLimitedAssortativity}, the 

%\begin{figure}[!t]
%	\centering
%    \includegraphics[width=0.45\textwidth]{figs/monoplex_unass_vs_multiplex_unass.eps}
%    \caption{
%        The comparison between monoplex networks and multiplex networks with limited assortativity.
%        We set $z = \lambda_{\text{blue}} = \lambda_{\text{red}}$.
%        }
%	\label{fig:expmonoplexVSMultiplexLimitedAssortativity}
%\end{figure}
%\begin{figure}[!t]
%	\centering
%    \includegraphics[width=0.45\textwidth]{figs/monoplex_unass_vs_multiplex_unass1.eps}
%    \caption{
%        The comparison between monoplex networks and multiplex networks with limited assortativity.
%        We set $z = \lambda_{\text{blue}} = \lambda_{\text{red}}$.
%        }
%	\label{fig:expmonoplexVSMultiplexLimitedAssortativity}
%\end{figure}

\begin{figure}[!t]
    \centering
    \subfigure[]{\hspace{0cm} \includegraphics[width=0.45\textwidth] {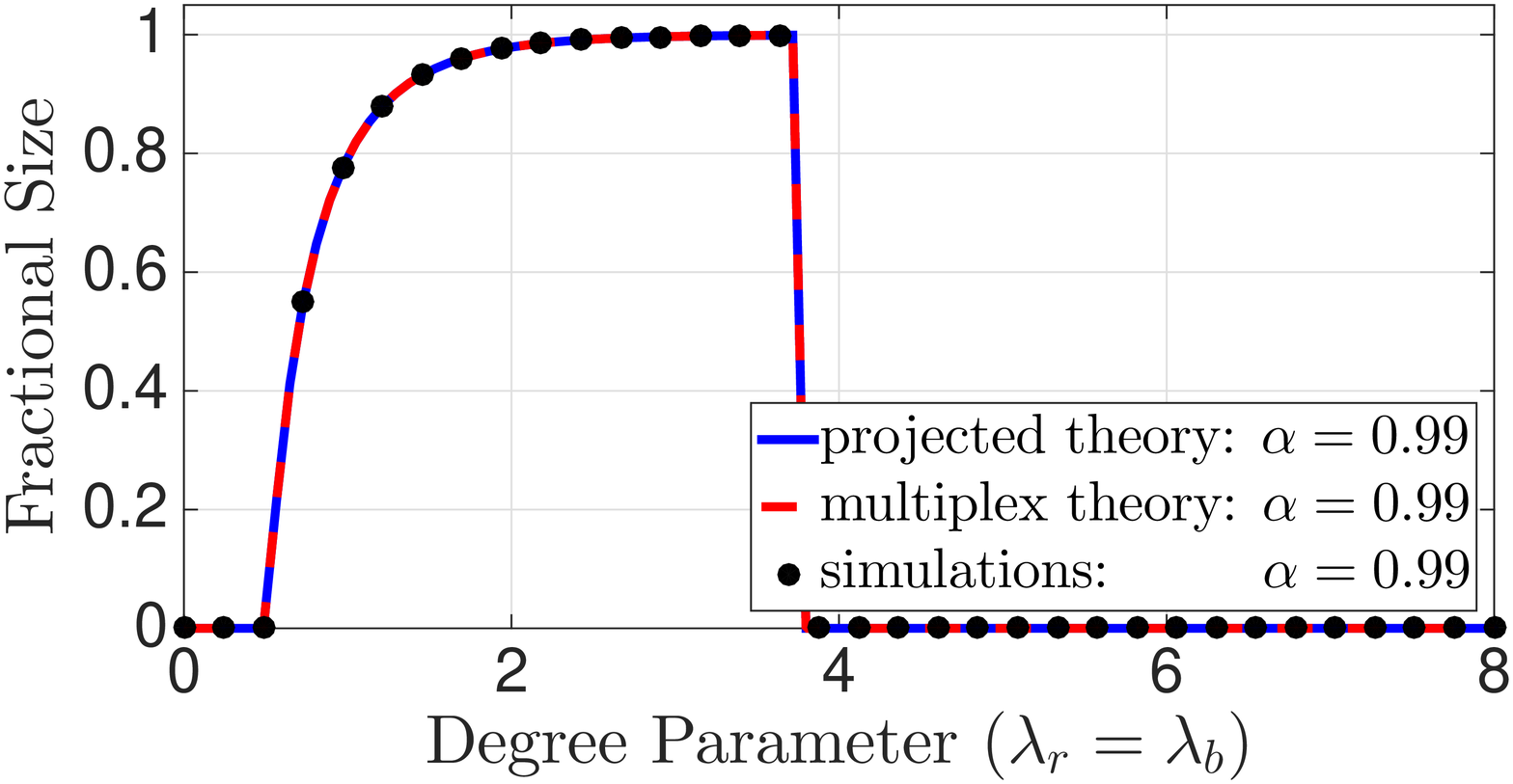}\label{fig:expmonoplexVSMultiplexLimitedAssortativity01}}
    \subfigure[]{\hspace{0cm} \includegraphics[width=0.45\textwidth] {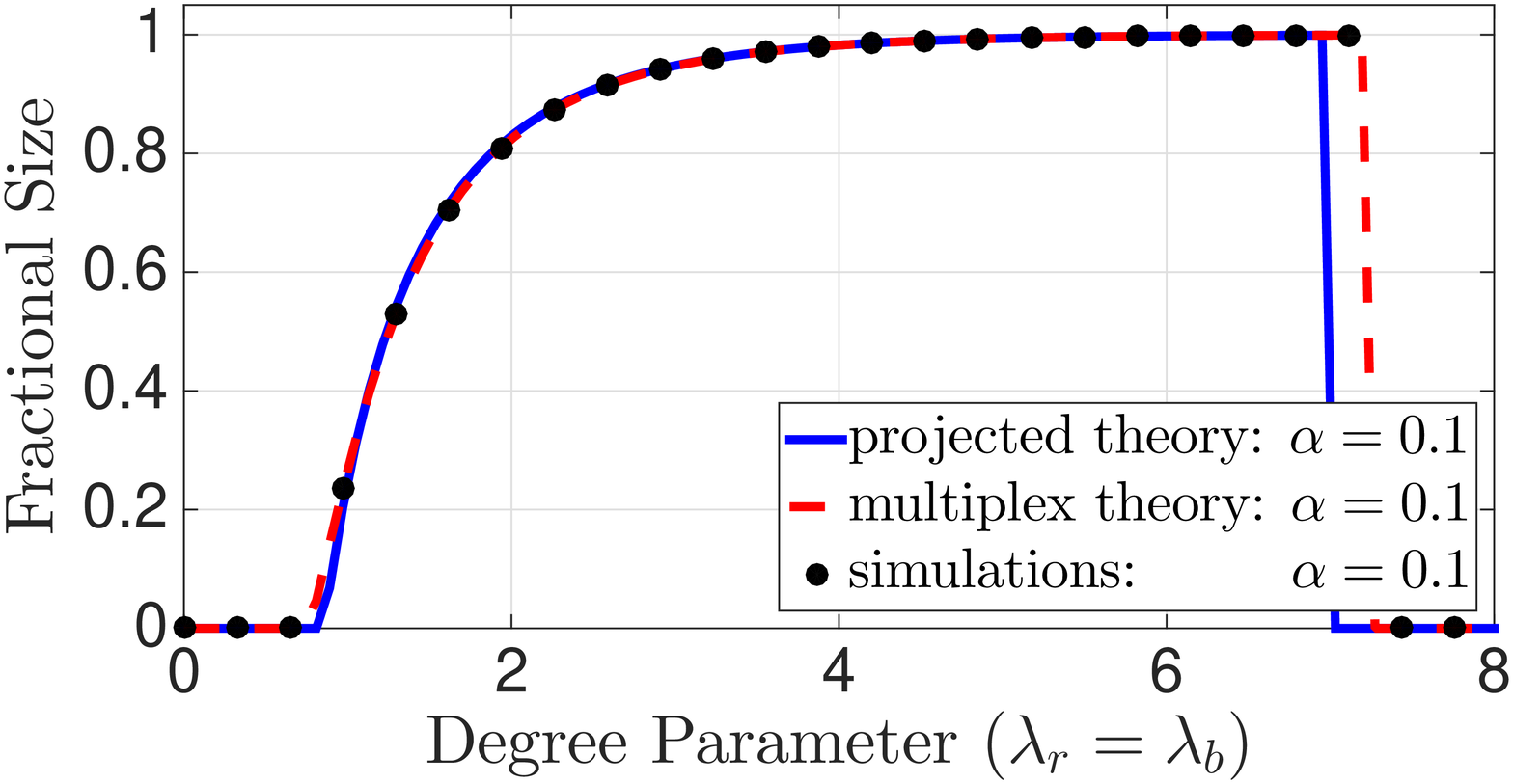}   \label{fig:expmonoplexVSMultiplexLimitedAssortativity099}}
    \caption{\sl Comparison between monoplex networks and multiplex networks with limited assortativity.
    In (a) and (b), we fix the threshold $\tau = 0.15$, the content parameter $c = 1$, then vary the degree parameters in (\ref{eq:degree_dist_colored_osy}).
    For the networks obtained by projected theory and the networks in multiplex theory with $\alpha = 0.99$, assortativity is negligible.
    However, when $\alpha = 0.1$, the assortativity coefficient of the networks in the multiplex theory become significant; e.g., it can be up to 0.21. 
    }
\label{fig:expmonoplexVSMultiplexLimitedAssortativity}
\end{figure}

\subsection{Multiplex Networks with Assortativity}
\label{sec:networkWithAssortativity}
In this section, we change the setting slightly to generate multiplex networks with high assortativity.
To that end, we use the degree distributions given at 
(\ref{eq:degree_dist_colored_osy}), but instead of setting $\lambda_r=\lambda_b$, we enforce
\begin{align}
\alpha\lambda_{r} = \lambda_{b}
\label{eq:lambda_rb_alfa}
\end{align}
for any $\alpha \in (0,1)$.     
This setting allows us to tune assortativity without changing the mean degree in the network. In particular, assortativity will increase as $\alpha$ decreases (by virtue of a small fraction of nodes forming a highly-connected cluster) \cite{zhuang2015information}. In addition, this setting allows us to compare the contagion dynamics in multi-layer networks when the upper layer is i) small but densely connected (small $\alpha$)
versus ii) large but loosely connected (large $\alpha$); see \cite{zhuang2015information} for relevant results for bond percolation processes.

%if we decrease $\alpha$, the corresponding level of assortativity will increase without changing average degree of the network, which has been demonstrated in .
%As we want to compare the complex contagions in multiplex networks with monoplex networks, we use the degree distribution (\ref{eq:degree_dist_colored_osy}) to create a monoplex network as the approach introduced in Section \ref{sec:networkWithLimitedAssortativity}.

\begin{figure}[!t]
    \centering
    \subfigure[]{\hspace{0cm} \includegraphics[totalheight=0.18\textheight,width=0.45\textwidth] {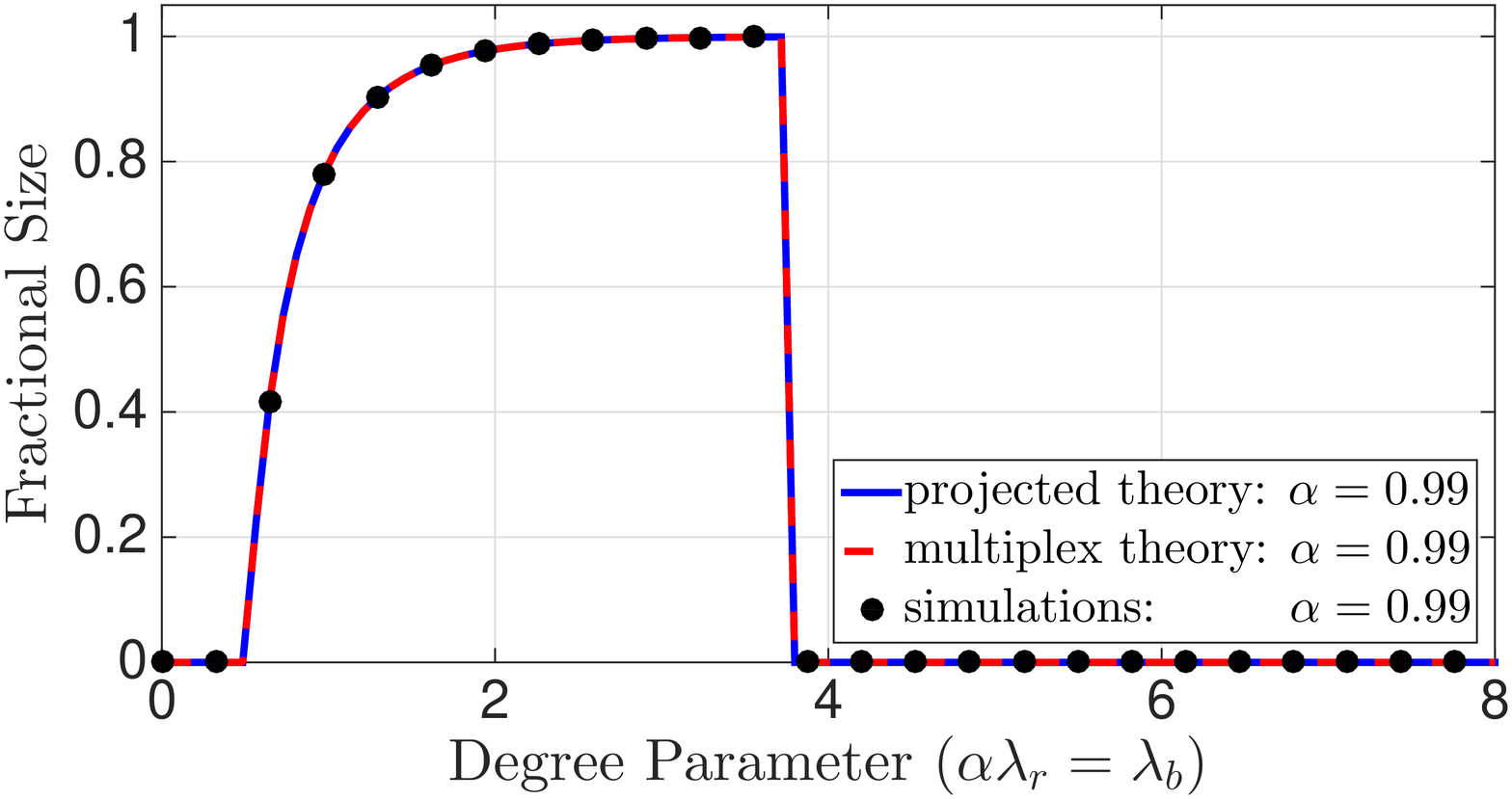}\label{fig:expmonoplexVSMultiplexAssortativity01}}
    \subfigure[]{\hspace{0cm} \includegraphics[totalheight=0.18\textheight,width=0.45\textwidth] {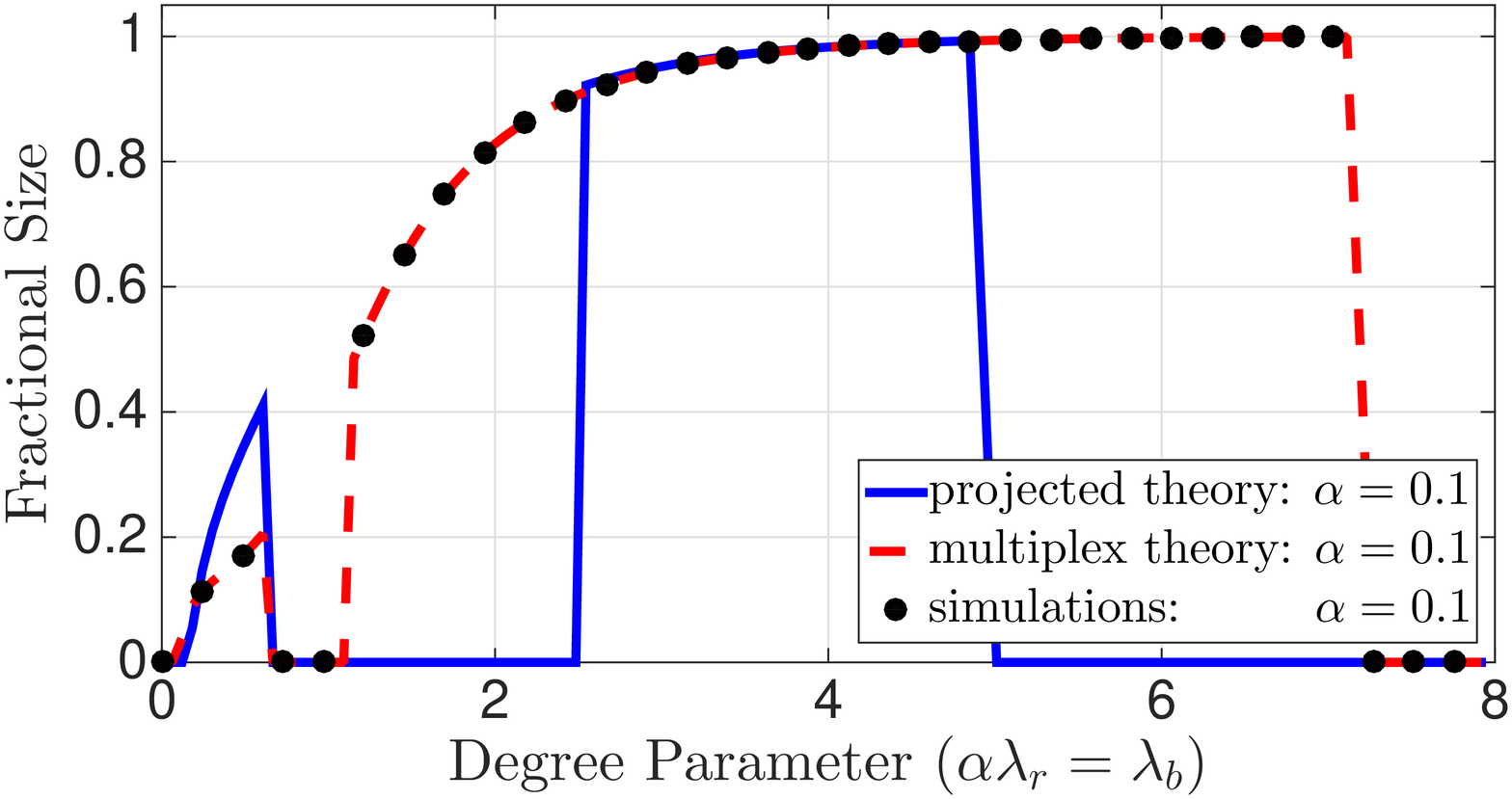} \label{fig:expmonoplexVSMultiplexAssortativity099}}
    \caption{\sl Comparison between monoplex networks and multiplex networks with assortativity.
    Similar with the observation in Figure \ref{fig:expmonoplexVSMultiplexLimitedAssortativity},  networks in the projected theory and in the multiplex theory with $\alpha = 0.99$ have negligible assortativity coefficients.
    However, for the networks of multiplex theory with $\alpha = 0.1$, assortativity coefficient ranges from $0.19$ to $0.79$.
    In general, assortativity increases  
    with increasing  $\lambda_{r}$ and $\lambda_{b}$ in the multiplex theory.
    }
\label{fig:expmonoplexVSMultiplexAssortativity}
\end{figure}

Using the above degree distributions, we generate monoplex and multiplex networks as in Section \ref{sec:networkWithLimitedAssortativity} and analyze the complex contagion process. 
In Figure \ref{fig:expmonoplexVSMultiplexAssortativity01}, we see that  when $\alpha = 0.99$, which leads to very limited assortativity, the difference between monoplex and multiplex networks is negligible. This is in parallel with what we observed in Section \ref{sec:networkWithLimitedAssortativity}.
However, decreasing $\alpha$ to 0.1 leads to two interesting observations 
in Figure \ref{fig:expmonoplexVSMultiplexAssortativity099}.
%not only the process of complex contagions in both of monoplex and multiplex networks changes, but the difference between monoplex and multiplex networks also becomes larger.
First, instead of the commonly reported two phase transitions  \cite{WattsExternal, OY13}, we observe four phase transitions in the cascade size as $\alpha\lambda_{r} = \lambda_{b}$ increases. Secondly we see a significant difference between the monoplex projected theory and multiplex theory, with multiplex theory matching the simulations perfectly. Once again, this shows that monoplex theory is unable to capture the cascade dynamics under certain settings.

The emergence of four phase transitions in Figure \ref{fig:expmonoplexVSMultiplexAssortativity099}, which to the best of our knowledge was {\em not} reported before, can be explained as follows.
When $\alpha=0.1$, only ten percent of the nodes have {\em red} edges, but the the mean number of red edges for those nodes equals $10 \lambda_b$ (see (\ref{eq:lambda_rb_alfa})). Therefore, the first couple of phase transitions taking place at very small $\lambda_b$ values can be attributed mainly to red-edges. First, $\lambda_b$ becomes large enough (e.g., gets to around 0.1) that the sub-network induced only on the red-edges contains a giant vulnerable cluster, giving rise to global cascades; note that at this point $\lambda_b$ is so small that blue edges do not create enough local stability to prevent cascades from happening. However, after $\lambda_b$ reaches a certain level (around $0.65$), the subgraph on red-edges, having average degree of $10 \lambda_b$, reaches the second phase transition point where  cascades stop due to the increased local connectivity of nodes. These first two transitions being second- and first-order, respectively, also confirms that they are primarily due to the red-edges. 

As $\lambda_b$ increases further we observe an interval where there are no global cascades due to either colors of edges; nodes with red {\em and} blue-edges are highly stubborn while nodes with only blue edges are not connected enough to trigger a cascade. This interval is then followed by a region where $\lambda_b$ is large enough that the sub-graph on blue edges has a giant vulnerable cluster. However, the emergence of a second-order transition in the whole network is prevented here due to some of these nodes turning stubborn as a result of their red-edges. Eventually, however, $\lambda_b$ becomes large enough that even with occasional stubborn nodes present, a giant vulnerable cluster emerges. This point is reached much later in monoplex networks as compared to the multiplex networks. This is because in the former case stubborn nodes (with red edges) are equally likely to be connected with any other node, while in the latter case they are mostly connected with each other; thus in the latter case they are less likely to inhibit the emergence of a giant vulnerable cluster on blue edges. 

Finally, the system goes through a fourth transition when $\lambda_b$ becomes large enough that even nodes with only blue edges become highly connected and hence stubborn. We see that this final transition point is reached much later in multiplex networks than monoplex networks meaning that cascades take place over a broader range of $\lambda_b$ values in the former case. Again, this can be attributed to the
high assortativity seen in multiplex networks that leads to {\em extremely} stubborn nodes (that have both blue and red edges) being isolated from those that are {\em mildly} stubborn (that have only blue edges). On the other hand, in monoplex networks,
every node is able to connect with the extremely stubborn nodes, and thus the critical value of $\lambda_b$ at which cascades become impossible due to high local stability is reached much earlier than that in multiplex networks.

\subsection{Two vs. Four Phase Transitions}
\label{sec:networkMultiplePhases}

In Section \ref{sec:networkWithAssortativity}, we have observed the possibility of having more than two phase transitions in the cascade size.  As discussed  there, multiple phase transitions occur mainly due to the setting (\ref{eq:lambda_rb_alfa})
that, with small $\alpha$, ensures a small fraction of  nodes having significantly higher connectivity than the rest, while also being mostly connected with each other.
Since the existence of more than two transitions has not been reported in previous studies, we are interested in exploring it further. In particular, 
we now investigate the impact of $\alpha$ on the number of phase transitions as well as transition points. 
Of particular interest will be to find the critical $\alpha$ value
that separates the cases where four phase transitions occur from those with only two transitions; e.g., the $\alpha$ value for which the two cascade regions overlap. 
For simplicity we only consider multiplex networks in this section.

%in the previous literature\cite{WattsExternal, OY13}.

%The reason why more than two phase transitions occur is that $\alpha^{*}$ controls when nodes with only blue edges and nodes with two colors of edges are able to have a global cascade.
%In the previous section, we have observed that when $\alpha^{*} = 0.1$ the two regions separate with each other while they fully overlap when $\alpha^{*} = 0.99$.

\begin{figure}[!t]
	\centering
    \includegraphics[width=0.5\textwidth]{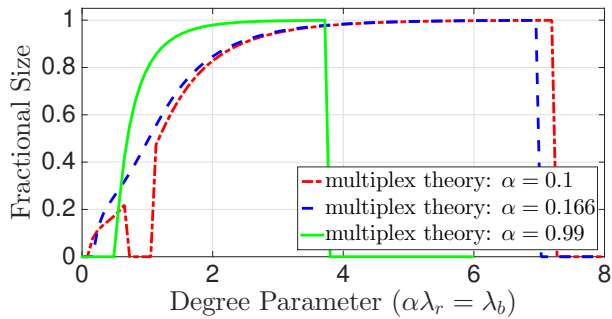}
    \caption{\sl
        Demonstration of multiple phase transitions.
        %We keep $\lambda_{b} = \alpha\lambda_{r}$.
        }
	\label{fig:expMultiplexPhase}
\end{figure}

%which is like what Hackett et al. \cite{hackett2015bond} have observed in bond percolation on multiplex networks.
Figure \ref{fig:expMultiplexPhase} shows the expected size of global cascades under (\ref{eq:degree_dist_colored_osy})-(\ref{eq:lambda_rb_alfa})
for three different values of $\alpha$. We see that 
global cascades take place over a single interval of $\alpha\lambda_r = \lambda_b$ when $\alpha$ is {\em large} (e.g., $\alpha=0.99$) while
over two disjoint intervals when $\alpha$ is {\em small} (e.g., $\alpha=0.1$). When $\alpha$ is somewhere in between (e.g., case $\alpha= 0.166$ ) it is possible to have the cascade intervals partially overlap. In such cases, we only see a single interval where global cascades take place. However, an additional transition point appears, manifested by a shift of slope in cascade size, marking possibly the overlapping point of (what would be) the two cascade intervals. 

Figure \ref{fig:expMultiplexPhase}  allows us to comment also on the impact of the size and density of the overlaying social network in facilitating influence propagation. With (\ref{eq:lambda_rb_alfa}) in effect, a small $\alpha$ corresponds to a social network with few but densely connected individuals, while large $\alpha$ corresponds to a social network with many subscribers, each with few connections on average. In all cases, the total number of edges in the social network is fixed by virtue of (\ref{eq:lambda_rb_alfa}). 
We see from Figure \ref{fig:expMultiplexPhase} that the comparison between the three cases leads to a multi-faceted picture as the mean number of links $\alpha \lambda_r$ varies.
For instance, the large but loosely connected case of $\alpha = 0.99$ leads to the largest expected global cascade size over a certain interval, but it has the smallest {\em cascade interval} among all three. The intermediate case of $\alpha= 0.166$ seems like a stretched version (over the $x$-axis) of the case with $\alpha=0.99$. In particular, this case leads to the largest interval where global cascades are possible, though the expected cascade size is smaller than that obtained with $\alpha = 0.99$ (and also with $\alpha=0.1)$ over certain intervals. Finally, the case of a small but densely connected overlay (i.e., with $\alpha=0.1$), falls right under the case with $\alpha=0.166$ for most values of $\alpha \lambda_r$, though it gives the largest size of all three in small intervals where $\alpha \lambda_r$ is very small or very large.

%\begin{table}[!h]
%    \centering
%    \begin{tabular}{c|c|r|r|}
%        Network Types    &   $\alpha$ &   Assortativity   &   Clust. Coefficients\\
%       \hline
%       \multirow{2}{*}{monoplex$\dagger$} &   0.1 &   xxx&  xxx    \\
%                                &   0.99    &   0.xxx&  0.xxx    \\
%                                           \hline
%       \multirow{2}{*}{Multiplex$\dagger$}   &   0.1 &   0.xxx&  0.xxx    \\
%                                    &   0.99    &   0.xxx&  0.xxx    \\
%    \end{tabular}
%    \caption{Statistics corresponding to the monoplex and multiplex networks}
%    \label{table:statisticsCompareCC}
%\end{table}

\section{Conclusion and Future Work}

\label{sec:conclusion}

We studied the diffusion of influence in clustered multiplex networks. 
We  solved analytically for the condition, probability, and expected size of {\em global} cascades,  and confirmed our results 
 via extensive computer simulations.
One of our key findings is to show how clustering affects the probability and expected size of global cascades.
%To the best of our knowledge, this works is the first work on the influence diffusion in clustered multiplex networks.
 We also compared several interesting properties of complex contagions on a multiplex network and its monoplex projection.
We demonstrate that  ignoring link types and aggregating network layers may lead to inaccurate conclusions about contagion dynamics, particularly when assortativity is high. Finally, we show for the first time that linear threshold models do not necessarily exhibit two phase transitions as previously reported. Depending on assortativity, we show that both in monoplex and multiplex cases (with two link types) it is possible to observe four phase transitions.

Our analysis and modeling framework subsumes some previous studies. For instance, by setting $h_{rt}(x) = h_{bt}(x) = 1$ in the recursive relations, we ensure that $\br$ and $\bb$ are non-clustered random networks.
So, our analysis corresponds to complex contagions in non-clustered networks, which was studied in \cite{YaganPRE}.
Similarly, if we let $h_{rs}(x) = h_{rt}(x) = 1$ in the recursions and set the content parameter $c= 1$, then our analysis corresponds to {\em complex} contagions in {\em clustered monoplex} networks, which was studied in \cite{hackett2011cascades}. 

%Besides, there are still some interesting open problems.
Future work may consider in more details the impact of assortativity or other topological features on the cascade dynamics. It would also be interesting to compare multiplex networks and their monoplex projections in terms of other dynamical processes; e.g., site percolation, transport processes, etc. 

%instead of cycles with size of three.

\section*{Acknowledgment}
This research was supported in part by National Science Foundation through grant CCF \#1422165, and in part by the Department of Electrical and Computer Engineering at Carnegie Mellon University. A.A. acknowledge financial support by the Spanish government through grant FIS2015-38266, ICREA Academia and James S. McDonnell Foundation.
\bibliography{./sdp}

\begin{thebibliography}{62}%
\makeatletter
\providecommand \@ifxundefined [1]{%
 \@ifx{#1\undefined}
}%
\providecommand \@ifnum [1]{%
 \ifnum #1\expandafter \@firstoftwo
 \else \expandafter \@secondoftwo
 \fi
}%
\providecommand \@ifx [1]{%
 \ifx #1\expandafter \@firstoftwo
 \else \expandafter \@secondoftwo
 \fi
}%
\providecommand \natexlab [1]{#1}%
\providecommand \enquote  [1]{``#1''}%
\providecommand \bibnamefont  [1]{#1}%
\providecommand \bibfnamefont [1]{#1}%
\providecommand \citenamefont [1]{#1}%
\providecommand \href@noop [0]{\@secondoftwo}%
\providecommand \href [0]{\begingroup \@sanitize@url \@href}%
\providecommand \@href[1]{\@@startlink{#1}\@@href}%
\providecommand \@@href[1]{\endgroup#1\@@endlink}%
\providecommand \@sanitize@url [0]{\catcode `\\12\catcode `\$12\catcode
  `\&12\catcode `\#12\catcode `\^12\catcode `\_12\catcode `\%12\relax}%
\providecommand \@@startlink[1]{}%
\providecommand \@@endlink[0]{}%
\providecommand \url  [0]{\begingroup\@sanitize@url \@url }%
\providecommand \@url [1]{\endgroup\@href {#1}{\urlprefix }}%
\providecommand \urlprefix  [0]{URL }%
\providecommand \Eprint [0]{\href }%
\providecommand \doibase [0]{http://dx.doi.org/}%
\providecommand \selectlanguage [0]{\@gobble}%
\providecommand \bibinfo  [0]{\@secondoftwo}%
\providecommand \bibfield  [0]{\@secondoftwo}%
\providecommand \translation [1]{[#1]}%
\providecommand \BibitemOpen [0]{}%
\providecommand \bibitemStop [0]{}%
\providecommand \bibitemNoStop [0]{.\EOS\space}%
\providecommand \EOS [0]{\spacefactor3000\relax}%
\providecommand \BibitemShut  [1]{\csname bibitem#1\endcsname}%
\let\auto@bib@innerbib\@empty
%</preamble>
\bibitem [{\citenamefont {Buldyrev}\ \emph {et~al.}(2010)\citenamefont
  {Buldyrev}, \citenamefont {Parshani}, \citenamefont {Paul}, \citenamefont
  {Stanley},\ and\ \citenamefont {Havlin}}]{Buldyrev}%
  \BibitemOpen
  \bibfield  {author} {\bibinfo {author} {\bibfnamefont {S.~V.}\ \bibnamefont
  {Buldyrev}}, \bibinfo {author} {\bibfnamefont {R.}~\bibnamefont {Parshani}},
  \bibinfo {author} {\bibfnamefont {G.}~\bibnamefont {Paul}}, \bibinfo {author}
  {\bibfnamefont {H.~E.}\ \bibnamefont {Stanley}}, \ and\ \bibinfo {author}
  {\bibfnamefont {S.}~\bibnamefont {Havlin}},\ }\href@noop {} {\bibfield
  {journal} {\bibinfo  {journal} {Nature}\ }\textbf {\bibinfo {volume} {464}},\
  \bibinfo {pages} {1025} (\bibinfo {year} {2010})}\BibitemShut {NoStop}%
\bibitem [{\citenamefont {Vespignani}(2010)}]{Vespignani}%
  \BibitemOpen
  \bibfield  {author} {\bibinfo {author} {\bibfnamefont {A.}~\bibnamefont
  {Vespignani}},\ }\href@noop {} {\bibfield  {journal} {\bibinfo  {journal}
  {Nature}\ }\textbf {\bibinfo {volume} {464}},\ \bibinfo {pages} {984}
  (\bibinfo {year} {2010})}\BibitemShut {NoStop}%
\bibitem [{\citenamefont {Ya\u{g}an}\ \emph {et~al.}(2012)\citenamefont
  {Ya\u{g}an}, \citenamefont {Qian}, \citenamefont {Zhang},\ and\ \citenamefont
  {Cochran}}]{YaganQianZhangCochranLong}%
  \BibitemOpen
  \bibfield  {author} {\bibinfo {author} {\bibfnamefont {O.}~\bibnamefont
  {Ya\u{g}an}}, \bibinfo {author} {\bibfnamefont {D.}~\bibnamefont {Qian}},
  \bibinfo {author} {\bibfnamefont {J.}~\bibnamefont {Zhang}}, \ and\ \bibinfo
  {author} {\bibfnamefont {D.}~\bibnamefont {Cochran}},\ }\href@noop {}
  {\bibfield  {journal} {\bibinfo  {journal} {IEEE Transactions on Parallel and
  Distributed Systems}\ }\textbf {\bibinfo {volume} {23}},\ \bibinfo {pages}
  {1708} (\bibinfo {year} {2012})},\ \bibinfo {note} {%Also available online at
  arXiv:1201.2698v2.}\BibitemShut {Stop}%
\bibitem [{\citenamefont {Brummitt}\ \emph
  {et~al.}(2012{\natexlab{a}})\citenamefont {Brummitt}, \citenamefont
  {D'Souza},\ and\ \citenamefont {Leicht}}]{brummitt2012suppressing}%
  \BibitemOpen
  \bibfield  {author} {\bibinfo {author} {\bibfnamefont {C.~D.}\ \bibnamefont
  {Brummitt}}, \bibinfo {author} {\bibfnamefont {R.~M.}\ \bibnamefont
  {D'Souza}}, \ and\ \bibinfo {author} {\bibfnamefont {E.}~\bibnamefont
  {Leicht}},\ }\href@noop {} {\bibfield  {journal} {\bibinfo  {journal}
  {Proceedings of the National Academy of Sciences}\ }\textbf {\bibinfo
  {volume} {109}},\ \bibinfo {pages} {E680} (\bibinfo {year}
  {2012}{\natexlab{a}})}\BibitemShut {NoStop}%
\bibitem [{\citenamefont {Radicchi}\ and\ \citenamefont
  {Arenas}(2013)}]{radicchi2013abrupt}%
  \BibitemOpen
  \bibfield  {author} {\bibinfo {author} {\bibfnamefont {F.}~\bibnamefont
  {Radicchi}}\ and\ \bibinfo {author} {\bibfnamefont {A.}~\bibnamefont
  {Arenas}},\ }\href@noop {} {\bibfield  {journal} {\bibinfo  {journal} {Nature
  Physics}\ }\textbf {\bibinfo {volume} {9}},\ \bibinfo {pages} {717} (\bibinfo
  {year} {2013})}\BibitemShut {NoStop}%
\bibitem [{\citenamefont {Reis}\ \emph {et~al.}(2014)\citenamefont {Reis},
  \citenamefont {Hu}, \citenamefont {Babino}, \citenamefont {Andrade~Jr},
  \citenamefont {Canals}, \citenamefont {Sigman},\ and\ \citenamefont
  {Makse}}]{reis2014avoiding}%
  \BibitemOpen
  \bibfield  {author} {\bibinfo {author} {\bibfnamefont {S.~D.}\ \bibnamefont
  {Reis}}, \bibinfo {author} {\bibfnamefont {Y.}~\bibnamefont {Hu}}, \bibinfo
  {author} {\bibfnamefont {A.}~\bibnamefont {Babino}}, \bibinfo {author}
  {\bibfnamefont {J.~S.}\ \bibnamefont {Andrade~Jr}}, \bibinfo {author}
  {\bibfnamefont {S.}~\bibnamefont {Canals}}, \bibinfo {author} {\bibfnamefont
  {M.}~\bibnamefont {Sigman}}, \ and\ \bibinfo {author} {\bibfnamefont {H.~A.}\
  \bibnamefont {Makse}},\ }\href@noop {} {\bibfield  {journal} {\bibinfo
  {journal} {Nature Physics}\ }\textbf {\bibinfo {volume} {10}},\ \bibinfo
  {pages} {762} (\bibinfo {year} {2014})}\BibitemShut {NoStop}%
\bibitem [{\citenamefont {Albert}\ \emph {et~al.}(2000)\citenamefont {Albert},
  \citenamefont {Jeong},\ and\ \citenamefont {Barab{\'a}si}}]{albert2000error}%
  \BibitemOpen
  \bibfield  {author} {\bibinfo {author} {\bibfnamefont {R.}~\bibnamefont
  {Albert}}, \bibinfo {author} {\bibfnamefont {H.}~\bibnamefont {Jeong}}, \
  and\ \bibinfo {author} {\bibfnamefont {A.-L.}\ \bibnamefont {Barab{\'a}si}},\
  }\href@noop {} {\bibfield  {journal} {\bibinfo  {journal} {nature}\ }\textbf
  {\bibinfo {volume} {406}},\ \bibinfo {pages} {378} (\bibinfo {year}
  {2000})}\BibitemShut {NoStop}%
\bibitem [{\citenamefont {Cohen}\ \emph {et~al.}(2000)\citenamefont {Cohen},
  \citenamefont {Erez}, \citenamefont {Ben-Avraham},\ and\ \citenamefont
  {Havlin}}]{cohen2000resilience}%
  \BibitemOpen
  \bibfield  {author} {\bibinfo {author} {\bibfnamefont {R.}~\bibnamefont
  {Cohen}}, \bibinfo {author} {\bibfnamefont {K.}~\bibnamefont {Erez}},
  \bibinfo {author} {\bibfnamefont {D.}~\bibnamefont {Ben-Avraham}}, \ and\
  \bibinfo {author} {\bibfnamefont {S.}~\bibnamefont {Havlin}},\ }\href@noop {}
  {\bibfield  {journal} {\bibinfo  {journal} {Physical review letters}\
  }\textbf {\bibinfo {volume} {85}},\ \bibinfo {pages} {4626} (\bibinfo {year}
  {2000})}\BibitemShut {NoStop}%
\bibitem [{\citenamefont {Cohen}\ \emph {et~al.}(2002)\citenamefont {Cohen},
  \citenamefont {Ben-Avraham},\ and\ \citenamefont
  {Havlin}}]{cohen2002percolation}%
  \BibitemOpen
  \bibfield  {author} {\bibinfo {author} {\bibfnamefont {R.}~\bibnamefont
  {Cohen}}, \bibinfo {author} {\bibfnamefont {D.}~\bibnamefont {Ben-Avraham}},
  \ and\ \bibinfo {author} {\bibfnamefont {S.}~\bibnamefont {Havlin}},\
  }\href@noop {} {\bibfield  {journal} {\bibinfo  {journal} {Physical Review
  E}\ }\textbf {\bibinfo {volume} {66}},\ \bibinfo {pages} {036113} (\bibinfo
  {year} {2002})}\BibitemShut {NoStop}%
\bibitem [{\citenamefont {Leicht}\ and\ \citenamefont
  {D'Souza}(2009)}]{leicht2009percolation}%
  \BibitemOpen
  \bibfield  {author} {\bibinfo {author} {\bibfnamefont {E.}~\bibnamefont
  {Leicht}}\ and\ \bibinfo {author} {\bibfnamefont {R.~M.}\ \bibnamefont
  {D'Souza}},\ }\href@noop {} {\bibfield  {journal} {\bibinfo  {journal} {arXiv
  preprint arXiv:0907.0894}\ } (\bibinfo {year} {2009})}\BibitemShut {NoStop}%
\bibitem [{\citenamefont {Nagler}\ \emph {et~al.}(2011)\citenamefont {Nagler},
  \citenamefont {Levina},\ and\ \citenamefont {Timme}}]{nagler2011impact}%
  \BibitemOpen
  \bibfield  {author} {\bibinfo {author} {\bibfnamefont {J.}~\bibnamefont
  {Nagler}}, \bibinfo {author} {\bibfnamefont {A.}~\bibnamefont {Levina}}, \
  and\ \bibinfo {author} {\bibfnamefont {M.}~\bibnamefont {Timme}},\
  }\href@noop {} {\bibfield  {journal} {\bibinfo  {journal} {Nature Physics}\
  }\textbf {\bibinfo {volume} {7}},\ \bibinfo {pages} {265} (\bibinfo {year}
  {2011})}\BibitemShut {NoStop}%
\bibitem [{\citenamefont {Murray}(2002)}]{Murray}%
  \BibitemOpen
  \bibfield  {author} {\bibinfo {author} {\bibfnamefont {J.~D.}\ \bibnamefont
  {Murray}},\ }\href@noop {} {\emph {\bibinfo {title} {Mathematical
  Biology}}},\ \bibinfo {edition} {3rd}\ ed.\ (\bibinfo  {publisher}
  {Springer},\ \bibinfo {address} {New York (NY)},\ \bibinfo {year}
  {2002})\BibitemShut {NoStop}%
\bibitem [{\citenamefont {Son}\ \emph {et~al.}(2011)\citenamefont {Son},
  \citenamefont {Grassberger},\ and\ \citenamefont
  {Paczuski}}]{son2011percolation}%
  \BibitemOpen
  \bibfield  {author} {\bibinfo {author} {\bibfnamefont {S.-W.}\ \bibnamefont
  {Son}}, \bibinfo {author} {\bibfnamefont {P.}~\bibnamefont {Grassberger}}, \
  and\ \bibinfo {author} {\bibfnamefont {M.}~\bibnamefont {Paczuski}},\
  }\href@noop {} {\bibfield  {journal} {\bibinfo  {journal} {Physical review
  letters}\ }\textbf {\bibinfo {volume} {107}},\ \bibinfo {pages} {195702}
  (\bibinfo {year} {2011})}\BibitemShut {NoStop}%
\bibitem [{\citenamefont {Baxter}\ \emph {et~al.}(2012)\citenamefont {Baxter},
  \citenamefont {Dorogovtsev}, \citenamefont {Goltsev},\ and\ \citenamefont
  {Mendes}}]{baxter2012avalanche}%
  \BibitemOpen
  \bibfield  {author} {\bibinfo {author} {\bibfnamefont {G.}~\bibnamefont
  {Baxter}}, \bibinfo {author} {\bibfnamefont {S.}~\bibnamefont {Dorogovtsev}},
  \bibinfo {author} {\bibfnamefont {A.}~\bibnamefont {Goltsev}}, \ and\
  \bibinfo {author} {\bibfnamefont {J.}~\bibnamefont {Mendes}},\ }\href@noop {}
  {\bibfield  {journal} {\bibinfo  {journal} {Physical review letters}\
  }\textbf {\bibinfo {volume} {109}},\ \bibinfo {pages} {248701} (\bibinfo
  {year} {2012})}\BibitemShut {NoStop}%
\bibitem [{\citenamefont {Dickison}\ \emph {et~al.}(2012)\citenamefont
  {Dickison}, \citenamefont {Havlin},\ and\ \citenamefont
  {Stanley}}]{dickison2012epidemics}%
  \BibitemOpen
  \bibfield  {author} {\bibinfo {author} {\bibfnamefont {M.}~\bibnamefont
  {Dickison}}, \bibinfo {author} {\bibfnamefont {S.}~\bibnamefont {Havlin}}, \
  and\ \bibinfo {author} {\bibfnamefont {H.~E.}\ \bibnamefont {Stanley}},\
  }\href@noop {} {\bibfield  {journal} {\bibinfo  {journal} {Physical Review
  E}\ }\textbf {\bibinfo {volume} {85}},\ \bibinfo {pages} {066109} (\bibinfo
  {year} {2012})}\BibitemShut {NoStop}%
\bibitem [{\citenamefont {Anderson}\ and\ \citenamefont
  {May}(1991)}]{AndersonMay}%
  \BibitemOpen
  \bibfield  {author} {\bibinfo {author} {\bibfnamefont {R.~M.}\ \bibnamefont
  {Anderson}}\ and\ \bibinfo {author} {\bibfnamefont {R.~M.}\ \bibnamefont
  {May}},\ }\href@noop {} {\emph {\bibinfo {title} {Infectious Diseases of
  Humans}}}\ (\bibinfo  {publisher} {Oxford University Press},\ \bibinfo
  {address} {Oxford (UK)},\ \bibinfo {year} {1991})\BibitemShut {NoStop}%
\bibitem [{\citenamefont {Sahneh}\ \emph {et~al.}(2013)\citenamefont {Sahneh},
  \citenamefont {Scoglio},\ and\ \citenamefont
  {Van~Mieghem}}]{SahnehScoglioMieghem}%
  \BibitemOpen
  \bibfield  {author} {\bibinfo {author} {\bibfnamefont {F.}~\bibnamefont
  {Sahneh}}, \bibinfo {author} {\bibfnamefont {C.}~\bibnamefont {Scoglio}}, \
  and\ \bibinfo {author} {\bibfnamefont {P.}~\bibnamefont {Van~Mieghem}},\
  }\href {\doibase 10.1109/TNET.2013.2239658} {\bibfield  {journal} {\bibinfo
  {journal} {Networking, IEEE/ACM Transactions on}\ }\textbf {\bibinfo {volume}
  {21}},\ \bibinfo {pages} {1609} (\bibinfo {year} {2013})}\BibitemShut
  {NoStop}%
\bibitem [{\citenamefont {Ya\u{g}an}\ \emph {et~al.}(2013)\citenamefont
  {Ya\u{g}an}, \citenamefont {Qian}, \citenamefont {Zhang},\ and\ \citenamefont
  {Cochran}}]{OY13}%
  \BibitemOpen
  \bibfield  {author} {\bibinfo {author} {\bibfnamefont {O.}~\bibnamefont
  {Ya\u{g}an}}, \bibinfo {author} {\bibfnamefont {D.}~\bibnamefont {Qian}},
  \bibinfo {author} {\bibfnamefont {J.}~\bibnamefont {Zhang}}, \ and\ \bibinfo
  {author} {\bibfnamefont {D.}~\bibnamefont {Cochran}},\ }\href@noop {}
  {\bibfield  {journal} {\bibinfo  {journal} {{IEEE} Journal on Selected Areas
  in Communications}\ }\textbf {\bibinfo {volume} {31}},\ \bibinfo {pages}
  {1038} (\bibinfo {year} {2013})}\BibitemShut {NoStop}%
\bibitem [{\citenamefont {Valdez}\ \emph {et~al.}(2013)\citenamefont {Valdez},
  \citenamefont {Macri}, \citenamefont {Stanley},\ and\ \citenamefont
  {Braunstein}}]{valdez2013triple}%
  \BibitemOpen
  \bibfield  {author} {\bibinfo {author} {\bibfnamefont {L.~D.}\ \bibnamefont
  {Valdez}}, \bibinfo {author} {\bibfnamefont {P.~A.}\ \bibnamefont {Macri}},
  \bibinfo {author} {\bibfnamefont {H.~E.}\ \bibnamefont {Stanley}}, \ and\
  \bibinfo {author} {\bibfnamefont {L.~A.}\ \bibnamefont {Braunstein}},\
  }\href@noop {} {\bibfield  {journal} {\bibinfo  {journal} {Physical Review
  E}\ }\textbf {\bibinfo {volume} {88}},\ \bibinfo {pages} {050803} (\bibinfo
  {year} {2013})}\BibitemShut {NoStop}%
\bibitem [{\citenamefont {Zhao}\ and\ \citenamefont
  {Bianconi}(2013)}]{zhao2013percolation}%
  \BibitemOpen
  \bibfield  {author} {\bibinfo {author} {\bibfnamefont {K.}~\bibnamefont
  {Zhao}}\ and\ \bibinfo {author} {\bibfnamefont {G.}~\bibnamefont
  {Bianconi}},\ }\href@noop {} {\bibfield  {journal} {\bibinfo  {journal}
  {Journal of Statistical Mechanics: Theory and Experiment}\ }\textbf {\bibinfo
  {volume} {2013}},\ \bibinfo {pages} {P05005} (\bibinfo {year}
  {2013})}\BibitemShut {NoStop}%
\bibitem [{\citenamefont {Cellai}\ \emph {et~al.}(2013)\citenamefont {Cellai},
  \citenamefont {L{\'o}pez}, \citenamefont {Zhou}, \citenamefont {Gleeson},\
  and\ \citenamefont {Bianconi}}]{cellai2013percolation}%
  \BibitemOpen
  \bibfield  {author} {\bibinfo {author} {\bibfnamefont {D.}~\bibnamefont
  {Cellai}}, \bibinfo {author} {\bibfnamefont {E.}~\bibnamefont {L{\'o}pez}},
  \bibinfo {author} {\bibfnamefont {J.}~\bibnamefont {Zhou}}, \bibinfo {author}
  {\bibfnamefont {J.~P.}\ \bibnamefont {Gleeson}}, \ and\ \bibinfo {author}
  {\bibfnamefont {G.}~\bibnamefont {Bianconi}},\ }\href@noop {} {\bibfield
  {journal} {\bibinfo  {journal} {Physical Review E}\ }\textbf {\bibinfo
  {volume} {88}},\ \bibinfo {pages} {052811} (\bibinfo {year}
  {2013})}\BibitemShut {NoStop}%
\bibitem [{\citenamefont {Azimi-Tafreshi}\ \emph {et~al.}(2014)\citenamefont
  {Azimi-Tafreshi}, \citenamefont {G{\'o}mez-Garde{\~n}es},\ and\ \citenamefont
  {Dorogovtsev}}]{azimi2014k}%
  \BibitemOpen
  \bibfield  {author} {\bibinfo {author} {\bibfnamefont {N.}~\bibnamefont
  {Azimi-Tafreshi}}, \bibinfo {author} {\bibfnamefont {J.}~\bibnamefont
  {G{\'o}mez-Garde{\~n}es}}, \ and\ \bibinfo {author} {\bibfnamefont
  {S.}~\bibnamefont {Dorogovtsev}},\ }\href@noop {} {\bibfield  {journal}
  {\bibinfo  {journal} {Physical Review E}\ }\textbf {\bibinfo {volume} {90}},\
  \bibinfo {pages} {032816} (\bibinfo {year} {2014})}\BibitemShut {NoStop}%
\bibitem [{\citenamefont {Bianconi}\ and\ \citenamefont
  {Dorogovtsev}(2014)}]{bianconi2014multiple}%
  \BibitemOpen
  \bibfield  {author} {\bibinfo {author} {\bibfnamefont {G.}~\bibnamefont
  {Bianconi}}\ and\ \bibinfo {author} {\bibfnamefont {S.~N.}\ \bibnamefont
  {Dorogovtsev}},\ }\href@noop {} {\bibfield  {journal} {\bibinfo  {journal}
  {Physical Review E}\ }\textbf {\bibinfo {volume} {89}},\ \bibinfo {pages}
  {062814} (\bibinfo {year} {2014})}\BibitemShut {NoStop}%
\bibitem [{\citenamefont {Radicchi}(2015)}]{radicchi2015percolation}%
  \BibitemOpen
  \bibfield  {author} {\bibinfo {author} {\bibfnamefont {F.}~\bibnamefont
  {Radicchi}},\ }\href@noop {} {\bibfield  {journal} {\bibinfo  {journal}
  {Nature Physics}\ }\textbf {\bibinfo {volume} {11}},\ \bibinfo {pages} {597}
  (\bibinfo {year} {2015})}\BibitemShut {NoStop}%
\bibitem [{\citenamefont {Gleeson}(2008)}]{gleeson2008cascades}%
  \BibitemOpen
  \bibfield  {author} {\bibinfo {author} {\bibfnamefont {J.~P.}\ \bibnamefont
  {Gleeson}},\ }\href@noop {} {\bibfield  {journal} {\bibinfo  {journal}
  {Physical Review E}\ }\textbf {\bibinfo {volume} {77}},\ \bibinfo {pages}
  {046117} (\bibinfo {year} {2008})}\BibitemShut {NoStop}%
\bibitem [{\citenamefont {Watts}(2002)}]{WattsExternal}%
  \BibitemOpen
  \bibfield  {author} {\bibinfo {author} {\bibfnamefont {D.~J.}\ \bibnamefont
  {Watts}},\ }\href@noop {} {\bibfield  {journal} {\bibinfo  {journal}
  {Proceedings of the National Academy of Sciences}\ }\textbf {\bibinfo
  {volume} {99}},\ \bibinfo {pages} {5766} (\bibinfo {year}
  {2002})}\BibitemShut {NoStop}%
\bibitem [{\citenamefont {Ya\u{g}an}\ and\ \citenamefont
  {Gligor}(2012)}]{YaganPRE}%
  \BibitemOpen
  \bibfield  {author} {\bibinfo {author} {\bibfnamefont {O.}~\bibnamefont
  {Ya\u{g}an}}\ and\ \bibinfo {author} {\bibfnamefont {V.}~\bibnamefont
  {Gligor}},\ }\href {\doibase 10.1103/PhysRevE.86.036103} {\bibfield
  {journal} {\bibinfo  {journal} {Physical Review E}\ }\textbf {\bibinfo
  {volume} {86}},\ \bibinfo {pages} {036103} (\bibinfo {year}
  {2012})}\BibitemShut {NoStop}%
\bibitem [{\citenamefont {Hackett}\ \emph {et~al.}(2016)\citenamefont
  {Hackett}, \citenamefont {Cellai}, \citenamefont {G\'omez}, \citenamefont
  {Arenas},\ and\ \citenamefont {Gleeson}}]{Alex_PRX}%
  \BibitemOpen
  \bibfield  {author} {\bibinfo {author} {\bibfnamefont {A.}~\bibnamefont
  {Hackett}}, \bibinfo {author} {\bibfnamefont {D.}~\bibnamefont {Cellai}},
  \bibinfo {author} {\bibfnamefont {S.}~\bibnamefont {G\'omez}}, \bibinfo
  {author} {\bibfnamefont {A.}~\bibnamefont {Arenas}}, \ and\ \bibinfo {author}
  {\bibfnamefont {J.~P.}\ \bibnamefont {Gleeson}},\ }\href {\doibase
  10.1103/PhysRevX.6.021002} {\bibfield  {journal} {\bibinfo  {journal} {Phys.
  Rev. X}\ }\textbf {\bibinfo {volume} {6}},\ \bibinfo {pages} {021002}
  (\bibinfo {year} {2016})}\BibitemShut {NoStop}%
\bibitem [{\citenamefont {Zhuang}\ and\ \citenamefont
  {Ya\u{g}an}(2016)}]{zhuang2015information}%
  \BibitemOpen
  \bibfield  {author} {\bibinfo {author} {\bibfnamefont {Y.}~\bibnamefont
  {Zhuang}}\ and\ \bibinfo {author} {\bibfnamefont {O.}~\bibnamefont
  {Ya\u{g}an}},\ }\href {\doibase 10.1109/TNSE.2016.2600059} {\bibfield
  {journal} {\bibinfo  {journal} {IEEE Transactions on Network Science and
  Engineering}\ }\textbf {\bibinfo {volume} {PP}},\ \bibinfo {pages} {1}
  (\bibinfo {year} {2016})}\BibitemShut {NoStop}%
\bibitem [{\citenamefont {Brummitt}\ \emph
  {et~al.}(2012{\natexlab{b}})\citenamefont {Brummitt}, \citenamefont {Lee},\
  and\ \citenamefont {Goh}}]{brummitt2012multiplexity}%
  \BibitemOpen
  \bibfield  {author} {\bibinfo {author} {\bibfnamefont {C.~D.}\ \bibnamefont
  {Brummitt}}, \bibinfo {author} {\bibfnamefont {K.-M.}\ \bibnamefont {Lee}}, \
  and\ \bibinfo {author} {\bibfnamefont {K.-I.}\ \bibnamefont {Goh}},\
  }\href@noop {} {\bibfield  {journal} {\bibinfo  {journal} {Physical Review
  E}\ }\textbf {\bibinfo {volume} {85}},\ \bibinfo {pages} {045102} (\bibinfo
  {year} {2012}{\natexlab{b}})}\BibitemShut {NoStop}%
\bibitem [{\citenamefont {Kivel{\"a}}\ \emph {et~al.}(2014)\citenamefont
  {Kivel{\"a}}, \citenamefont {Arenas}, \citenamefont {Barthelemy},
  \citenamefont {Gleeson}, \citenamefont {Moreno},\ and\ \citenamefont
  {Porter}}]{kivela2014multilayer}%
  \BibitemOpen
  \bibfield  {author} {\bibinfo {author} {\bibfnamefont {M.}~\bibnamefont
  {Kivel{\"a}}}, \bibinfo {author} {\bibfnamefont {A.}~\bibnamefont {Arenas}},
  \bibinfo {author} {\bibfnamefont {M.}~\bibnamefont {Barthelemy}}, \bibinfo
  {author} {\bibfnamefont {J.~P.}\ \bibnamefont {Gleeson}}, \bibinfo {author}
  {\bibfnamefont {Y.}~\bibnamefont {Moreno}}, \ and\ \bibinfo {author}
  {\bibfnamefont {M.~A.}\ \bibnamefont {Porter}},\ }\href@noop {} {\bibfield
  {journal} {\bibinfo  {journal} {Journal of complex networks}\ }\textbf
  {\bibinfo {volume} {2}},\ \bibinfo {pages} {203} (\bibinfo {year}
  {2014})}\BibitemShut {NoStop}%
\bibitem [{\citenamefont {De~Domenico}\ \emph {et~al.}(2014)\citenamefont
  {De~Domenico}, \citenamefont {Sol\'e-Ribalta}, \citenamefont {G\'omez},\ and\
  \citenamefont {Arenas}}]{Domenico10062014}%
  \BibitemOpen
  \bibfield  {author} {\bibinfo {author} {\bibfnamefont {M.}~\bibnamefont
  {De~Domenico}}, \bibinfo {author} {\bibfnamefont {A.}~\bibnamefont
  {Sol\'e-Ribalta}}, \bibinfo {author} {\bibfnamefont {S.}~\bibnamefont
  {G\'omez}}, \ and\ \bibinfo {author} {\bibfnamefont {A.}~\bibnamefont
  {Arenas}},\ }\href {\doibase 10.1073/pnas.1318469111} {\ \textbf {\bibinfo
  {volume} {111}},\ \bibinfo {pages} {8351} (\bibinfo {year}
  {2014})}\BibitemShut {NoStop}%
\bibitem [{\citenamefont {De~Domenico}\ \emph {et~al.}(2013)\citenamefont
  {De~Domenico}, \citenamefont {Sol{\'e}-Ribalta}, \citenamefont {Cozzo},
  \citenamefont {Kivel{\"a}}, \citenamefont {Moreno}, \citenamefont {Porter},
  \citenamefont {G{\'o}mez},\ and\ \citenamefont
  {Arenas}}]{de2013mathematical}%
  \BibitemOpen
  \bibfield  {author} {\bibinfo {author} {\bibfnamefont {M.}~\bibnamefont
  {De~Domenico}}, \bibinfo {author} {\bibfnamefont {A.}~\bibnamefont
  {Sol{\'e}-Ribalta}}, \bibinfo {author} {\bibfnamefont {E.}~\bibnamefont
  {Cozzo}}, \bibinfo {author} {\bibfnamefont {M.}~\bibnamefont {Kivel{\"a}}},
  \bibinfo {author} {\bibfnamefont {Y.}~\bibnamefont {Moreno}}, \bibinfo
  {author} {\bibfnamefont {M.~A.}\ \bibnamefont {Porter}}, \bibinfo {author}
  {\bibfnamefont {S.}~\bibnamefont {G{\'o}mez}}, \ and\ \bibinfo {author}
  {\bibfnamefont {A.}~\bibnamefont {Arenas}},\ }\href@noop {} {\bibfield
  {journal} {\bibinfo  {journal} {Physical Review X}\ }\textbf {\bibinfo
  {volume} {3}},\ \bibinfo {pages} {041022} (\bibinfo {year}
  {2013})}\BibitemShut {NoStop}%
\bibitem [{\citenamefont {Min}\ \emph {et~al.}(2014)\citenamefont {Min},
  \citenamefont {Do~Yi}, \citenamefont {Lee},\ and\ \citenamefont
  {Goh}}]{min2014network}%
  \BibitemOpen
  \bibfield  {author} {\bibinfo {author} {\bibfnamefont {B.}~\bibnamefont
  {Min}}, \bibinfo {author} {\bibfnamefont {S.}~\bibnamefont {Do~Yi}}, \bibinfo
  {author} {\bibfnamefont {K.-M.}\ \bibnamefont {Lee}}, \ and\ \bibinfo
  {author} {\bibfnamefont {K.-I.}\ \bibnamefont {Goh}},\ }\href@noop {}
  {\bibfield  {journal} {\bibinfo  {journal} {Physical Review E}\ }\textbf
  {\bibinfo {volume} {89}},\ \bibinfo {pages} {042811} (\bibinfo {year}
  {2014})}\BibitemShut {NoStop}%
\bibitem [{\citenamefont {Serrano}\ \emph {et~al.}(2015)\citenamefont
  {Serrano}, \citenamefont {Buzna},\ and\ \citenamefont
  {Bogu{\~n}{\'a}}}]{serrano2015escaping}%
  \BibitemOpen
  \bibfield  {author} {\bibinfo {author} {\bibfnamefont {M.~{\'A}.}\
  \bibnamefont {Serrano}}, \bibinfo {author} {\bibfnamefont {L.}~\bibnamefont
  {Buzna}}, \ and\ \bibinfo {author} {\bibfnamefont {M.}~\bibnamefont
  {Bogu{\~n}{\'a}}},\ }\href@noop {} {\bibfield  {journal} {\bibinfo  {journal}
  {New Journal of Physics}\ }\textbf {\bibinfo {volume} {17}},\ \bibinfo
  {pages} {053033} (\bibinfo {year} {2015})}\BibitemShut {NoStop}%
\bibitem [{\citenamefont {Granell}\ \emph {et~al.}(2013)\citenamefont
  {Granell}, \citenamefont {G{\'o}mez},\ and\ \citenamefont
  {Arenas}}]{granell2013dynamical}%
  \BibitemOpen
  \bibfield  {author} {\bibinfo {author} {\bibfnamefont {C.}~\bibnamefont
  {Granell}}, \bibinfo {author} {\bibfnamefont {S.}~\bibnamefont {G{\'o}mez}},
  \ and\ \bibinfo {author} {\bibfnamefont {A.}~\bibnamefont {Arenas}},\
  }\href@noop {} {\bibfield  {journal} {\bibinfo  {journal} {Physical review
  letters}\ }\textbf {\bibinfo {volume} {111}},\ \bibinfo {pages} {128701}
  (\bibinfo {year} {2013})}\BibitemShut {NoStop}%
\bibitem [{\citenamefont {Baxter}\ \emph {et~al.}(2014)\citenamefont {Baxter},
  \citenamefont {Dorogovtsev}, \citenamefont {Mendes},\ and\ \citenamefont
  {Cellai}}]{baxter2014weak}%
  \BibitemOpen
  \bibfield  {author} {\bibinfo {author} {\bibfnamefont {G.~J.}\ \bibnamefont
  {Baxter}}, \bibinfo {author} {\bibfnamefont {S.~N.}\ \bibnamefont
  {Dorogovtsev}}, \bibinfo {author} {\bibfnamefont {J.~F.~F.}\ \bibnamefont
  {Mendes}}, \ and\ \bibinfo {author} {\bibfnamefont {D.}~\bibnamefont
  {Cellai}},\ }\href@noop {} {\bibfield  {journal} {\bibinfo  {journal}
  {Physical Review E}\ }\textbf {\bibinfo {volume} {89}},\ \bibinfo {pages}
  {042801} (\bibinfo {year} {2014})}\BibitemShut {NoStop}%
\bibitem [{\citenamefont {De~Domenico}\ \emph {et~al.}(2016)\citenamefont
  {De~Domenico}, \citenamefont {Granell}, \citenamefont {Porter},\ and\
  \citenamefont {Arenas}}]{de2016physics}%
  \BibitemOpen
  \bibfield  {author} {\bibinfo {author} {\bibfnamefont {M.}~\bibnamefont
  {De~Domenico}}, \bibinfo {author} {\bibfnamefont {C.}~\bibnamefont
  {Granell}}, \bibinfo {author} {\bibfnamefont {M.~A.}\ \bibnamefont {Porter}},
  \ and\ \bibinfo {author} {\bibfnamefont {A.}~\bibnamefont {Arenas}},\
  }\href@noop {} {\bibfield  {journal} {\bibinfo  {journal} {arXiv preprint
  arXiv:1604.02021}\ } (\bibinfo {year} {2016})}\BibitemShut {NoStop}%
\bibitem [{\citenamefont {Wu}\ \emph {et~al.}(2016)\citenamefont {Wu},
  \citenamefont {Arenas},\ and\ \citenamefont {G{\'o}mez}}]{wu2016influence}%
  \BibitemOpen
  \bibfield  {author} {\bibinfo {author} {\bibfnamefont {H.}~\bibnamefont
  {Wu}}, \bibinfo {author} {\bibfnamefont {A.}~\bibnamefont {Arenas}}, \ and\
  \bibinfo {author} {\bibfnamefont {S.}~\bibnamefont {G{\'o}mez}},\ }\href@noop
  {} {\bibfield  {journal} {\bibinfo  {journal} {arXiv preprint
  arXiv:1606.01688}\ } (\bibinfo {year} {2016})}\BibitemShut {NoStop}%
\bibitem [{\citenamefont {Boccaletti}\ \emph {et~al.}(2014)\citenamefont
  {Boccaletti}, \citenamefont {Bianconi}, \citenamefont {Criado}, \citenamefont
  {Del~Genio}, \citenamefont {G{\'o}mez-Garde{\~n}es}, \citenamefont {Romance},
  \citenamefont {Sendi{\~n}a-Nadal}, \citenamefont {Wang},\ and\ \citenamefont
  {Zanin}}]{boccaletti2014structure}%
  \BibitemOpen
  \bibfield  {author} {\bibinfo {author} {\bibfnamefont {S.}~\bibnamefont
  {Boccaletti}}, \bibinfo {author} {\bibfnamefont {G.}~\bibnamefont
  {Bianconi}}, \bibinfo {author} {\bibfnamefont {R.}~\bibnamefont {Criado}},
  \bibinfo {author} {\bibfnamefont {C.~I.}\ \bibnamefont {Del~Genio}}, \bibinfo
  {author} {\bibfnamefont {J.}~\bibnamefont {G{\'o}mez-Garde{\~n}es}}, \bibinfo
  {author} {\bibfnamefont {M.}~\bibnamefont {Romance}}, \bibinfo {author}
  {\bibfnamefont {I.}~\bibnamefont {Sendi{\~n}a-Nadal}}, \bibinfo {author}
  {\bibfnamefont {Z.}~\bibnamefont {Wang}}, \ and\ \bibinfo {author}
  {\bibfnamefont {M.}~\bibnamefont {Zanin}},\ }\href@noop {} {\bibfield
  {journal} {\bibinfo  {journal} {Physics Reports}\ }\textbf {\bibinfo {volume}
  {544}},\ \bibinfo {pages} {1} (\bibinfo {year} {2014})}\BibitemShut {NoStop}%
\bibitem [{\citenamefont {Cellai}\ and\ \citenamefont
  {Bianconi}(2016)}]{cellai2016multiplex}%
  \BibitemOpen
  \bibfield  {author} {\bibinfo {author} {\bibfnamefont {D.}~\bibnamefont
  {Cellai}}\ and\ \bibinfo {author} {\bibfnamefont {G.}~\bibnamefont
  {Bianconi}},\ }\href@noop {} {\bibfield  {journal} {\bibinfo  {journal}
  {Physical Review E}\ }\textbf {\bibinfo {volume} {93}},\ \bibinfo {pages}
  {032302} (\bibinfo {year} {2016})}\BibitemShut {NoStop}%
\bibitem [{\citenamefont {Newman}\ \emph {et~al.}(2001)\citenamefont {Newman},
  \citenamefont {Strogatz},\ and\ \citenamefont {Watts}}]{MN01}%
  \BibitemOpen
  \bibfield  {author} {\bibinfo {author} {\bibfnamefont {M.~E.}\ \bibnamefont
  {Newman}}, \bibinfo {author} {\bibfnamefont {S.~H.}\ \bibnamefont
  {Strogatz}}, \ and\ \bibinfo {author} {\bibfnamefont {D.~J.}\ \bibnamefont
  {Watts}},\ }\href@noop {} {\bibfield  {journal} {\bibinfo  {journal}
  {Physical review E}\ }\textbf {\bibinfo {volume} {64}},\ \bibinfo {pages}
  {026118} (\bibinfo {year} {2001})}\BibitemShut {NoStop}%
\bibitem [{\citenamefont {Newman}(2003)}]{MN03}%
  \BibitemOpen
  \bibfield  {author} {\bibinfo {author} {\bibfnamefont {M.~E.}\ \bibnamefont
  {Newman}},\ }\href@noop {} {\bibfield  {journal} {\bibinfo  {journal} {SIAM
  review}\ }\textbf {\bibinfo {volume} {45}},\ \bibinfo {pages} {167} (\bibinfo
  {year} {2003})}\BibitemShut {NoStop}%
\bibitem [{\citenamefont {Watts}\ and\ \citenamefont
  {Strogatz}(1998)}]{WattsStrogatz}%
  \BibitemOpen
  \bibfield  {author} {\bibinfo {author} {\bibfnamefont {D.~J.}\ \bibnamefont
  {Watts}}\ and\ \bibinfo {author} {\bibfnamefont {S.~H.}\ \bibnamefont
  {Strogatz}},\ }\href@noop {} {\bibfield  {journal} {\bibinfo  {journal}
  {Nature}\ }\textbf {\bibinfo {volume} {393}},\ \bibinfo {pages} {440}
  (\bibinfo {year} {1998})}\BibitemShut {NoStop}%
\bibitem [{\citenamefont {Serrano}\ and\ \citenamefont
  {Bogu\~n\'a}(2006)}]{SerranoBoguna2}%
  \BibitemOpen
  \bibfield  {author} {\bibinfo {author} {\bibfnamefont {M.~A.}\ \bibnamefont
  {Serrano}}\ and\ \bibinfo {author} {\bibfnamefont {M.}~\bibnamefont
  {Bogu\~n\'a}},\ }\href {\doibase 10.1103/PhysRevE.74.056114} {\bibfield
  {journal} {\bibinfo  {journal} {Phys. Rev. E}\ }\textbf {\bibinfo {volume}
  {74}},\ \bibinfo {pages} {056114} (\bibinfo {year} {2006})}\BibitemShut
  {NoStop}%
\bibitem [{\citenamefont {Newman}(2009)}]{MN09}%
  \BibitemOpen
  \bibfield  {author} {\bibinfo {author} {\bibfnamefont {M.~E.~J.}\
  \bibnamefont {Newman}},\ }\href@noop {} {\bibfield  {journal} {\bibinfo
  {journal} {Physical review letters}\ }\textbf {\bibinfo {volume} {103}},\
  \bibinfo {pages} {058701} (\bibinfo {year} {2009})}\BibitemShut {NoStop}%
\bibitem [{\citenamefont {Gleeson}\ \emph {et~al.}(2010)\citenamefont
  {Gleeson}, \citenamefont {Melnik},\ and\ \citenamefont
  {Hackett}}]{gleeson2010clustering}%
  \BibitemOpen
  \bibfield  {author} {\bibinfo {author} {\bibfnamefont {J.~P.}\ \bibnamefont
  {Gleeson}}, \bibinfo {author} {\bibfnamefont {S.}~\bibnamefont {Melnik}}, \
  and\ \bibinfo {author} {\bibfnamefont {A.}~\bibnamefont {Hackett}},\
  }\href@noop {} {\bibfield  {journal} {\bibinfo  {journal} {Physical Review
  E}\ }\textbf {\bibinfo {volume} {81}},\ \bibinfo {pages} {066114} (\bibinfo
  {year} {2010})}\BibitemShut {NoStop}%
\bibitem [{\citenamefont {Hackett}\ \emph
  {et~al.}(2011{\natexlab{a}})\citenamefont {Hackett}, \citenamefont
  {Gleeson},\ and\ \citenamefont {Melnik}}]{hackett2011site}%
  \BibitemOpen
  \bibfield  {author} {\bibinfo {author} {\bibfnamefont {A.}~\bibnamefont
  {Hackett}}, \bibinfo {author} {\bibfnamefont {J.~P.}\ \bibnamefont
  {Gleeson}}, \ and\ \bibinfo {author} {\bibfnamefont {S.}~\bibnamefont
  {Melnik}},\ }\href@noop {} {\bibfield  {journal} {\bibinfo  {journal} {Int.
  J. Complex Systems in Science}\ }\textbf {\bibinfo {volume} {1}},\ \bibinfo
  {pages} {25} (\bibinfo {year} {2011}{\natexlab{a}})}\BibitemShut {NoStop}%
\bibitem [{\citenamefont {Hackett}\ \emph
  {et~al.}(2011{\natexlab{b}})\citenamefont {Hackett}, \citenamefont {Melnik},\
  and\ \citenamefont {Gleeson}}]{hackett2011cascades}%
  \BibitemOpen
  \bibfield  {author} {\bibinfo {author} {\bibfnamefont {A.}~\bibnamefont
  {Hackett}}, \bibinfo {author} {\bibfnamefont {S.}~\bibnamefont {Melnik}}, \
  and\ \bibinfo {author} {\bibfnamefont {J.~P.}\ \bibnamefont {Gleeson}},\
  }\href@noop {} {\bibfield  {journal} {\bibinfo  {journal} {Physical Review
  E}\ }\textbf {\bibinfo {volume} {83}},\ \bibinfo {pages} {056107} (\bibinfo
  {year} {2011}{\natexlab{b}})}\BibitemShut {NoStop}%
\bibitem [{\citenamefont {Huang}\ \emph {et~al.}(2013)\citenamefont {Huang},
  \citenamefont {Shao}, \citenamefont {Wang}, \citenamefont {Buldyrev},
  \citenamefont {Stanley},\ and\ \citenamefont {Havlin}}]{huang2013robustness}%
  \BibitemOpen
  \bibfield  {author} {\bibinfo {author} {\bibfnamefont {X.}~\bibnamefont
  {Huang}}, \bibinfo {author} {\bibfnamefont {S.}~\bibnamefont {Shao}},
  \bibinfo {author} {\bibfnamefont {H.}~\bibnamefont {Wang}}, \bibinfo {author}
  {\bibfnamefont {S.~V.}\ \bibnamefont {Buldyrev}}, \bibinfo {author}
  {\bibfnamefont {H.~E.}\ \bibnamefont {Stanley}}, \ and\ \bibinfo {author}
  {\bibfnamefont {S.}~\bibnamefont {Havlin}},\ }\href@noop {} {\bibfield
  {journal} {\bibinfo  {journal} {EPL (Europhysics Letters)}\ }\textbf
  {\bibinfo {volume} {101}},\ \bibinfo {pages} {18002} (\bibinfo {year}
  {2013})}\BibitemShut {NoStop}%
\bibitem [{\citenamefont {Colomer-de Sim{\'o}n}\ and\ \citenamefont
  {Bogu{\~n}{\'a}}(2014)}]{colomer2014double}%
  \BibitemOpen
  \bibfield  {author} {\bibinfo {author} {\bibfnamefont {P.}~\bibnamefont
  {Colomer-de Sim{\'o}n}}\ and\ \bibinfo {author} {\bibfnamefont
  {M.}~\bibnamefont {Bogu{\~n}{\'a}}},\ }\href@noop {} {\bibfield  {journal}
  {\bibinfo  {journal} {Physical Review X}\ }\textbf {\bibinfo {volume} {4}},\
  \bibinfo {pages} {041020} (\bibinfo {year} {2014})}\BibitemShut {NoStop}%
\bibitem [{\citenamefont {Cozzo}\ \emph {et~al.}(2015)\citenamefont {Cozzo},
  \citenamefont {Kivel{\"a}}, \citenamefont {De~Domenico}, \citenamefont
  {Sol{\'e}-Ribalta}, \citenamefont {Arenas}, \citenamefont {G{\'o}mez},
  \citenamefont {Porter},\ and\ \citenamefont {Moreno}}]{cozzo2015structure}%
  \BibitemOpen
  \bibfield  {author} {\bibinfo {author} {\bibfnamefont {E.}~\bibnamefont
  {Cozzo}}, \bibinfo {author} {\bibfnamefont {M.}~\bibnamefont {Kivel{\"a}}},
  \bibinfo {author} {\bibfnamefont {M.}~\bibnamefont {De~Domenico}}, \bibinfo
  {author} {\bibfnamefont {A.}~\bibnamefont {Sol{\'e}-Ribalta}}, \bibinfo
  {author} {\bibfnamefont {A.}~\bibnamefont {Arenas}}, \bibinfo {author}
  {\bibfnamefont {S.}~\bibnamefont {G{\'o}mez}}, \bibinfo {author}
  {\bibfnamefont {M.~A.}\ \bibnamefont {Porter}}, \ and\ \bibinfo {author}
  {\bibfnamefont {Y.}~\bibnamefont {Moreno}},\ }\href@noop {} {\bibfield
  {journal} {\bibinfo  {journal} {New Journal of Physics}\ }\textbf {\bibinfo
  {volume} {17}},\ \bibinfo {pages} {073029} (\bibinfo {year}
  {2015})}\BibitemShut {NoStop}%
\bibitem [{\citenamefont {Miller}(2009)}]{JM09}%
  \BibitemOpen
  \bibfield  {author} {\bibinfo {author} {\bibfnamefont {J.~C.}\ \bibnamefont
  {Miller}},\ }\href@noop {} {\bibfield  {journal} {\bibinfo  {journal}
  {Physical Review E}\ }\textbf {\bibinfo {volume} {80}},\ \bibinfo {pages}
  {020901} (\bibinfo {year} {2009})}\BibitemShut {NoStop}%
\bibitem [{\citenamefont {Molloy}\ and\ \citenamefont {Reed}(1995)}]{MM95}%
  \BibitemOpen
  \bibfield  {author} {\bibinfo {author} {\bibfnamefont {M.}~\bibnamefont
  {Molloy}}\ and\ \bibinfo {author} {\bibfnamefont {B.}~\bibnamefont {Reed}},\
  }\href@noop {} {\bibfield  {journal} {\bibinfo  {journal} {Random structures
  \& algorithms}\ }\textbf {\bibinfo {volume} {6}},\ \bibinfo {pages} {161}
  (\bibinfo {year} {1995})}\BibitemShut {NoStop}%
\bibitem [{\citenamefont {Tang}\ \emph {et~al.}(2011)\citenamefont {Tang},
  \citenamefont {Yuan}, \citenamefont {Mao}, \citenamefont {Li}, \citenamefont
  {Chen},\ and\ \citenamefont {Dai}}]{TangYuanMaoLiChenDai}%
  \BibitemOpen
  \bibfield  {author} {\bibinfo {author} {\bibfnamefont {S.}~\bibnamefont
  {Tang}}, \bibinfo {author} {\bibfnamefont {J.}~\bibnamefont {Yuan}}, \bibinfo
  {author} {\bibfnamefont {X.}~\bibnamefont {Mao}}, \bibinfo {author}
  {\bibfnamefont {X.-Y.}\ \bibnamefont {Li}}, \bibinfo {author} {\bibfnamefont
  {W.}~\bibnamefont {Chen}}, \ and\ \bibinfo {author} {\bibfnamefont
  {G.}~\bibnamefont {Dai}},\ }in\ \href {\doibase 10.1109/INFCOM.2011.5935046}
  {\emph {\bibinfo {booktitle} {Proceedings of IEEE INFOCOM 2011}}}\ (\bibinfo
  {year} {2011})\ pp.\ \bibinfo {pages} {2291 --2299}\BibitemShut {NoStop}%
\bibitem [{\citenamefont {Dodds}\ and\ \citenamefont
  {Watts}(2004)}]{DoddsWatts}%
  \BibitemOpen
  \bibfield  {author} {\bibinfo {author} {\bibfnamefont {P.}~\bibnamefont
  {Dodds}}\ and\ \bibinfo {author} {\bibfnamefont {D.~J.}\ \bibnamefont
  {Watts}},\ }\href@noop {} {\bibfield  {journal} {\bibinfo  {journal} {Phys.
  Rev. Lett.}\ }\textbf {\bibinfo {volume} {92}},\ \bibinfo {pages} {218701}
  (\bibinfo {year} {2004})}\BibitemShut {NoStop}%
\bibitem [{\citenamefont {Athreya}\ and\ \citenamefont
  {Ney}(2012)}]{athreya2012branching}%
  \BibitemOpen
  \bibfield  {author} {\bibinfo {author} {\bibfnamefont {K.~B.}\ \bibnamefont
  {Athreya}}\ and\ \bibinfo {author} {\bibfnamefont {P.~E.}\ \bibnamefont
  {Ney}},\ }\href@noop {} {\emph {\bibinfo {title} {Branching processes}}},\
  Vol.\ \bibinfo {volume} {196}\ (\bibinfo  {publisher} {Springer Science \&
  Business Media},\ \bibinfo {year} {2012})\BibitemShut {NoStop}%
\bibitem [{\citenamefont {Wilf}(2013)}]{HW13}%
  \BibitemOpen
  \bibfield  {author} {\bibinfo {author} {\bibfnamefont {H.~S.}\ \bibnamefont
  {Wilf}},\ }\href@noop {} {\emph {\bibinfo {title}
  {generatingfunctionology}}}\ (\bibinfo  {publisher} {Elsevier},\ \bibinfo
  {year} {2013})\BibitemShut {NoStop}%
\bibitem [{\citenamefont {Gleeson}\ and\ \citenamefont
  {Cahalane}(2007)}]{gleeson2007seed}%
  \BibitemOpen
  \bibfield  {author} {\bibinfo {author} {\bibfnamefont {J.~P.}\ \bibnamefont
  {Gleeson}}\ and\ \bibinfo {author} {\bibfnamefont {D.~J.}\ \bibnamefont
  {Cahalane}},\ }\href@noop {} {\bibfield  {journal} {\bibinfo  {journal}
  {Physical Review E}\ }\textbf {\bibinfo {volume} {75}},\ \bibinfo {pages}
  {056103} (\bibinfo {year} {2007})}\BibitemShut {NoStop}%
\bibitem [{\citenamefont {Newman}(2002)}]{newman2002assortative}%
  \BibitemOpen
  \bibfield  {author} {\bibinfo {author} {\bibfnamefont {M.~E.}\ \bibnamefont
  {Newman}},\ }\href@noop {} {\bibfield  {journal} {\bibinfo  {journal}
  {Physical review letters}\ }\textbf {\bibinfo {volume} {89}},\ \bibinfo
  {pages} {208701} (\bibinfo {year} {2002})}\BibitemShut {NoStop}%
\bibitem [{\citenamefont {S{\"o}derberg}(2003)}]{soderberg2003random}%
  \BibitemOpen
  \bibfield  {author} {\bibinfo {author} {\bibfnamefont {B.}~\bibnamefont
  {S{\"o}derberg}},\ }\href@noop {} {\bibfield  {journal} {\bibinfo  {journal}
  {Physical Review E}\ }\textbf {\bibinfo {volume} {68}},\ \bibinfo {pages}
  {015102} (\bibinfo {year} {2003})}\BibitemShut {NoStop}%
\bibitem [{\citenamefont {S{\"o}derberg}(2002)}]{soderberg2002general}%
  \BibitemOpen
  \bibfield  {author} {\bibinfo {author} {\bibfnamefont {B.}~\bibnamefont
  {S{\"o}derberg}},\ }\href@noop {} {\bibfield  {journal} {\bibinfo  {journal}
  {Physical review E}\ }\textbf {\bibinfo {volume} {66}},\ \bibinfo {pages}
  {066121} (\bibinfo {year} {2002})}\BibitemShut {NoStop}%
\end{thebibliography}%

\end{document}